\documentclass[fleqn,usenatbib]{mnras}

\usepackage{newtxtext}%,newtxmath}
\usepackage{mathptmx}
\usepackage[T1]{fontenc}
\usepackage{ae,aecompl}
\usepackage{hyperref}

\usepackage{graphicx}	% Including figure files
\usepackage{amsmath}	% Advanced maths commands
\usepackage{amssymb} % Extra maths symbols
\usepackage{footnote}
\usepackage{adjustbox}
\usepackage{lipsum}
\usepackage{mwe}
\usepackage{verbatim}
\usepackage{blindtext}
\usepackage{mathtools}
\usepackage{multirow}
\usepackage[group-separator={,},group-minimum-digits=4]{siunitx}
\usepackage{xcolor}

\title[UVIT observation of NGC 4590]{Study of UV bright sources in globular cluster NGC 4590 using Ultraviolet Imaging Telescope (UVIT) observations}

\author[Kumar et al.]{
Ranjan Kumar,$^{1}$\thanks{E-mail: ranjankmr488@gmail.com}
Ananta C. Pradhan,$^{1}$\thanks{E-mail: acp.phy@gmail.com}
M. Parthasarathy,$^{2}$
Sonika Piridi,$^{1}$
\newauthor Santi Cassisi,$^{3,4}$ 
Devendra K. Ojha,$^{5}$ 
Abhisek Mohapatra,$^{1,6}$
and Jayant Murthy $^{2}$ \\
$^{1}$Department of Physics and Astronomy, National Institute of Technology, Rourkela, Odisha - 769 008, India\\
$^{2}$Indian Institute of Astrophysics, Bangalore - 560 034, India\\
$^{3}$INAF - Astronomical Observatory of Abruzzo, Via M. Maggini, sn. 64100 Teramo, Italy.\\
$^{4}$INFN -  Sezione di Pisa, Universit\`a di Pisa, Largo Pontecorvo 3, 56127 Pisa, Italy.\\
$^{5}$Department of Astronomy and Astrophysics, Tata Institute of Fundamental Research (TIFR), Mumbai - 400 005, India\\
$^{6}$Inter-University Centre for Astronomy and Astrophysics, Ganeshkhind, Pune 411007, India\\
}

\date{Accepted 2022 February 7. Received 2022 February 3; in original form 2022 January 4}

\pubyear{2022}

\begin{document}
\label{firstpage}
\pagerange{\pageref{firstpage}--\pageref{lastpage}}
\maketitle

\begin{abstract}
We have studied ultraviolet (UV) bright sources in the Galactic globular cluster (GGC) NGC 4590 using Ultraviolet Imaging Telescope (UVIT) on-board the \mbox{{\em AstroSat}} satellite. Using UV-optical color-magnitude diagrams (CMDs), we have identified and characterized the sources of different evolutionary stages i.e., blue horizontal branch stars (BHBs), extremely blue horizontal branch stars (EHBs), blue straggler stars (BSs), variable stars, etc. We estimated effective temperature (T$_{\mathrm{eff}}$), gravity ($\log$(g)), luminosity (L$_{bol}$), and hence the radius (R) of these hot stars by fitting spectral energy distribution (SED) with the help of stellar atmosphere models. Two new far-UV (FUV) bright cluster member stars situated near the core of the cluster have been detected; one of them is an EHB star and the other one is either in its post-blue hook evolutionary phase or in white dwarf phase. The evolutionary status of all the hot stars, identified in the cluster, has been investigated by using various evolutionary models. We find the massive and younger BSs are concentrated at the center of the cluster whereas the older and less massive BSs are distributed though out the cluster. The BSs normalized radial distribution seems to be bi-modal with a minimum located at r$_{\mathrm{min}}$ = 4.3 r$_c$. We calculated A$^+$ parameter of the cluster which is obtained using cumulative normalized radial distribution of horizontal branch stars (HBs) and BSs. We measured this value up to half-mass radius of the cluster  to be $+ 0.13$, which indicates that NGC 4590 is one of the youngest clusters among dynamically intermediate age GGCs with a dynamical age of $0.423\pm0.096$ Gyr. 
\end{abstract}
%(A$^+_{\mathrm{rh}})$

%%%%%%%%%%%%%%%%%%%%%%%%%%%%%%%%%%%%%%%%%%%%%%%%%%

%%%%%%%%%%%%%%%%% BODY OF PAPER %%%%%%%%%%%%%%%%%%

\section{Introduction}   \label{sec:introduction}

Globular clusters (GCs) are moderately nearby objects in the observable sky where it is possible to resolve and observe the properties of individual stars of various evolutionary stages. The sources with effective temperature (T$_\mathrm{eff}$) higher than 7,000 K mostly contribute to the ultraviolet (UV) emission of the GCs \citep{Schiavon2012, Brown2016, Moehler2019}. These UV bright sources in GCs are blue horizontal branch stars (BHBs), extremely blue horizontal branch stars (EHBs), post asymptotic giant branch stars (pAGBs), blue hook (BHk) stars, blue straggler stars (BSs), etc. 

BHBs are core He burning stars (HeCBs) located around a well defined sequence in color-magnitude diagrams (CMDs) \citep{Barnard1900, Bruggencate1927, Catelan2009}. The distribution of BHBs in CMDs is mainly driven by the mass of their envelope: the lower is the envelope mass (M$_{\mathrm{env}}$), the hotter is the horizontal branch (HB) location of the star \citep{Kippenhahn1990, Cassisi2013}. They can evolve through the early-AGB and then to the pAGB stages, or climb up the AGB experiencing a number of thermal pulses depending on the residual envelope mass at the end of the core He burning stage \citep{Dorman1993, Moehler2019}. However, EHBs are produced as a result of strong mass loss in red giant branch (RGB) phase. They are also HeCBs but with very thin envelope mass \citep[M$_{\mathrm{env}}<0.02$$ \,M_\odot$, ][]{Dorman1993, Moehler2019}. They never reach the AGB/pAGB phase and evolve through AGB-manqu\'e phase. The BHk stars are a part of the EHB population and evolve into AGB-manqu\'e stars. An extreme fine tuning in the mass loss efficiency along the RGB stage or a specific evolutionary channel including, e.g., binary evolution is needed in order to explain the peculiar evolutionary properties of BHk \citep{Sweigart1997, Brown2001, Cassisi2003}.

GCs also host BSs which are core hydrogen burning stars and strong sources of UV emission. In the CMDs, these stars appear brighter and hotter than typical stars at the main sequence turn-off (MSTO) and therefore would be expected to be more massive and to appear as an anomalously young stellar component in the GCs. BSs are formed as a consequence of mass transfer/merger in binary systems and/or direct collision in dense stellar environments \citep{Sollima2008, Knigge2009, Geller2011}. Their evolution is similar to the single star evolution track. However, depending upon the environment during merger or collision, the mass and age of BSs are slightly different from single star evolution \citep{Sills2009, Sun2021}. Due to their extremely large effective temperature, a UV imaging study is a powerful tool to identify and trace the properties of BSs in GCs \citep{Ferraro2000, Raso2017, Ferraro2018, Sindhu2019, Sahu2019, Subramaniam2020, Kumar2021a}.

The Ultraviolet Imaging Telescope (UVIT) on-board the \mbox{{\em AstroSat}} satellite \citep{Tandon2017} has observed about 20 -- 30 GCs since its launch in 2015 and have provided various key science results for about 10 of them. The physical parameters such as ($\mathrm{T_{eff}}$), radius, luminosity, mass, etc., have been derived for the UV bright sources of various evolutionary stages (EHBs, BHBs, BSs, etc.) using UVIT and other available optical photometric observations \citep{Sahu2019, Singh2020, Kumar2020, Kumar2021a, Kumar2021, Prabhu2021, Rani2021}. The stellar evolutionary models such as BaSTI (old) and BaSTI-IAC (updated) \citep{Pietrinferni2004, Hidalgo2018, Pietrinferni2020}, FSPS models \citep{Conroy2009}, Padova isochrone sets \citep{Marigo2007}, $\mathrm{Y^2}$ stellar evolutionary tracks \citep{Lee2015}, etc., have well incorporated in deriving these parameters.  The UVIT study has enabled to find hot binary companion of BSs, such as white dwarfs (WDs), and EHBs in the dense clusters, NGC 1851 \citep{Singh2020} and NGC 5466 \citep{Sahu2019a}. Similarly, evolutionary status of hot stars such as EHBs, and post-HB (pAGB / AGB-manqu\'e) has been studied in detail for NGC 2808 by \citet{Prabhu2021}.The evidence of multiple populations among HB stars are also noticed in NGC 1851, NGC 2808, NGC 7492, and NGC 4147 \citep{Subramaniam2017, Jain2019,  Kumar2021a, Kumar2021, Rani2021}. In this paper, we present UVIT observations of NGC 4590 (M68) to trace and characterize its hot stellar populations. 

NGC 4590 ($\mathrm{\alpha_{J2000} = 12^h39^m27^s.98,\ \delta_{J2000} = -26^\circ44'38''.6}$) is a very metal-poor globular cluster with $\mathrm{[Fe/H]} = -2.42\pm0.14$ \citep{Schaeuble2015} situated at a distance of 10.40 kpc \citep[][]{Baumgardt2021} in the northern Galactic hemisphere. Its age is 11.2 Gyr \citep[][2010 edition]{Harris1996} and line of sight extinction $E(B-V)$ is 0.05 \citep{Schlafly2011}. It contains a significant number of RR Lyrae stars along with BHBs and RHBs \citep{Walker1994, Kains2015}. It has been extensively studied for multiple population of RGB and main-sequence (MS) stars \citep{Milone2017}. The cluster also shows a presence of light-elements peculiarities, a difference in He content and mass loss efficiencies between the first- and second- generation populations \citep{Milone2018, Tailo2020}. The atmospheric parameters such as effective temperature, surface gravity, and metal contents of five BHBs, nine RHBs and 11 RGBs in the cluster have been derived by \cite{Schaeuble2015} using a detailed high resolution spectroscopic study. They also found Na - O anti-correlation, constant Ca-abundance and mild r-process enrichment. Similarly, \cite{Behr2003} derived $\mathrm{T_{eff}}$, log(g), rotational velocity and chemical abundances of 11 BHBs of the cluster using HIRES spectrograph of the Keck I telescope. \cite{Zinn1972} detected one UV-bright star in the cluster which later was found to be a field star using proper motion values from \textit{Gaia} \citep{Bond2021}. As this cluster is a perfect candidate for the investigation of the second parameter problem \citep{Cassisi2013, Milone2014} in low metalicity regime and to explore UV-bright stars, we observed it in far-UV (FUV: 1300 $-$ 1800 \AA) and near-UV (NUV: 2000 $-$ 2800 \AA) imaging study with UVIT onboard \mbox{{\em AstroSat}} \citep{Tandon2017}.

We describe the observation details and data reduction procedures in \S\ref{sec:observation} and in \S\ref{sec:cmd}, we discuss about the CMDs and identification of the sources of different evolutionary stages. In \S\ref{sec:sed}, we derive stellar parameters of the sources in our sample using spectral energy distribution (SED) and in \S\ref{sec:evol_status}, we have discussed the evolutionary status of the hot stars present in the cluster. The dynamical age of the cluster using BSs distribution is discussed in \S\ref{sec:rad_BSs}. Finally, we provide the conclusions of our study in \S\ref{sec:conclusion}.
 
\section{Observation and Data Reduction}
\label{sec:observation}

We observed the cluster NGC 4590 with three FUV filters (BaF2, Sapphire, and Silica) and three NUV filters (NUVB15, NUVB13, and NUVB4) of UVIT in several orbits/frames. The observation details of NGC 4590 are given in \autoref{tab:observation}. The observed data set is reduced using a customized software package {\tt CCDLAB} \citep{Postma2017} with astrometry from the {\em Gaia} DR2 source catalog \citep{GaiaCatalog2018}. The point spread function (PSF) photometry is performed in {\tt IRAF}\footnote{\href{https://iraf-community.github.io/}{https://iraf-community.github.io/}} \citep{Tody1986, Tody1993} software package with {\tt DAOPHOT} routine given by \cite{Stetson1987}. The photometry details of various filters are given in \autoref{tab:observation}.

The extinction value obtained for the cluster using IRAS extinction map\footnote{\href{https://irsa.ipac.caltech.edu/applications/DUST/}{https://irsa.ipac.caltech.edu/applications/DUST/}} is E(B$-$V)=0.05 \citep{Schlafly2011}. We calculated the respective extinction values in UVIT filters using \citet{cardeli1989} extinction law  and then magnitudes of all the sources were corrected for extinction.

\citet{Schiavon2012} have presented a photometric analysis of UV bright stars in the cluster using {\em Galaxy Evolution Explorer (GALEX)} imaging surveys and provided FUV-NUV vs FUV CMDs only for the FUV bright stars within 100$''$ to 2000$''$ radial distance of the cluster but not for the stars in the inner region (radial distance $< 100''$ towards center of the cluster) due to low resolution ($\sim 5''$) of the {\em GALEX}. They were also not able to confirm the cluster membership of the detected sources due to unavailability of the kinematic information during that time. We cross-matched the \mbox{{\em GALEX}} observed sources with {\em Global Astrometric Interferometer for Astrophysics (Gaia)} EDR3 cluster membership catalog \citep{Vasiliev2021} to confirm their cluster membership. We found only 42 cluster member stars among the 327 sources presented by \citet{Schiavon2012}. We quantitatively compared the FUV bright sources detected by \mbox{{\em GALEX}} and UVIT within the UVIT field of View (30$'$). There are about 42 additional sources detected in the FUV filter of UVIT within 100$''<$ radius $< 900''$ and 100 additional sources near the center of the cluster ($10'' <$ radius $< 100''$). 
 
In this paper, we provide a detailed UV photometric analysis of the hot sources within $10''$ to $900''$ radial distance from the center of the cluster using six UVIT filters. We have used various stellar evolutionary tracks and isochrones to derive physical parameters of the cluster and also to trace the evolutionary status of the confirmed cluster member stars. 

\subsection{Optical counterparts and cluster membership} \label{sec:counterparts}

\citet{Nardiello2018} have provided a complete photometric catalog and membership probability of stars for 56 GCs using WFC3/UVIS and ACS/WFC surveys of {\em Hubble Space Telescope (HST)} in five filters. Similarly the photometric catalog along with the cluster membership of the sources for 150  GCs is provided by \citet{Vasiliev2021} using {\em Gaia} EDR3 survey. The cluster membership information provided by these two surveys is quite useful for separating field stars from cluster members. {\em HST} photometric surveys are mainly devoted to the central and denser regions of GGCs (2.7$'\times$2.7$'$) with a very high angular resolution of $\sim 0.2'' - 0.4''$, whereas {\em Gaia} has observed the entire cluster with an angular resolution of 0.4$''$.

For the cluster NGC 4590, we found that sources detected in NUV filters (NUVB13 and NUVB4) are over-crowded in the central region whereas sources detected in FUV filters are well resolved and sparse due to the quite less severe crowding conditions in the FUV images. So, we selected only FUV detected cluster members and their NUV counterparts in the central region of the cluster using {\em HST} photometric catalog of \citet{Nardiello2018}. We used \citet{Vasiliev2021} catalog for both FUV and NUV bright cluster members in outer parts of the cluster (i.e., outside the 2.7$'\times$2.7$'$ of the central region) as well as for the FUV undetected NUV bright sources at the central regime of the cluster.  We have given the total number of member stars of the cluster detected in each filter in \autoref{tab:observation}. The ground-based photometry of NGC 4590 is obtained based on the observations of the cluster from 1984 to 2018 \citep[see Table 2 of ][]{Stetson2019} with various ground-based telescopes. We have cross-matched UVIT observed cluster-members with this catalog to get their optical counterparts in order to show the optical, NUV-optical and FUV-optical CMDs for the entire cluster. We converted the Johnson-Cousins UBVRI magnitudes into ABmag system using the conversion relation given by \citet{Blanton2007}. All the photometric magnitudes used in this paper are in AB magnitude system.

\section{Color-Magnitude Diagrams}
\label{sec:cmd}

\begin{figure*}
    \centering
    \includegraphics[width=0.495\textwidth]{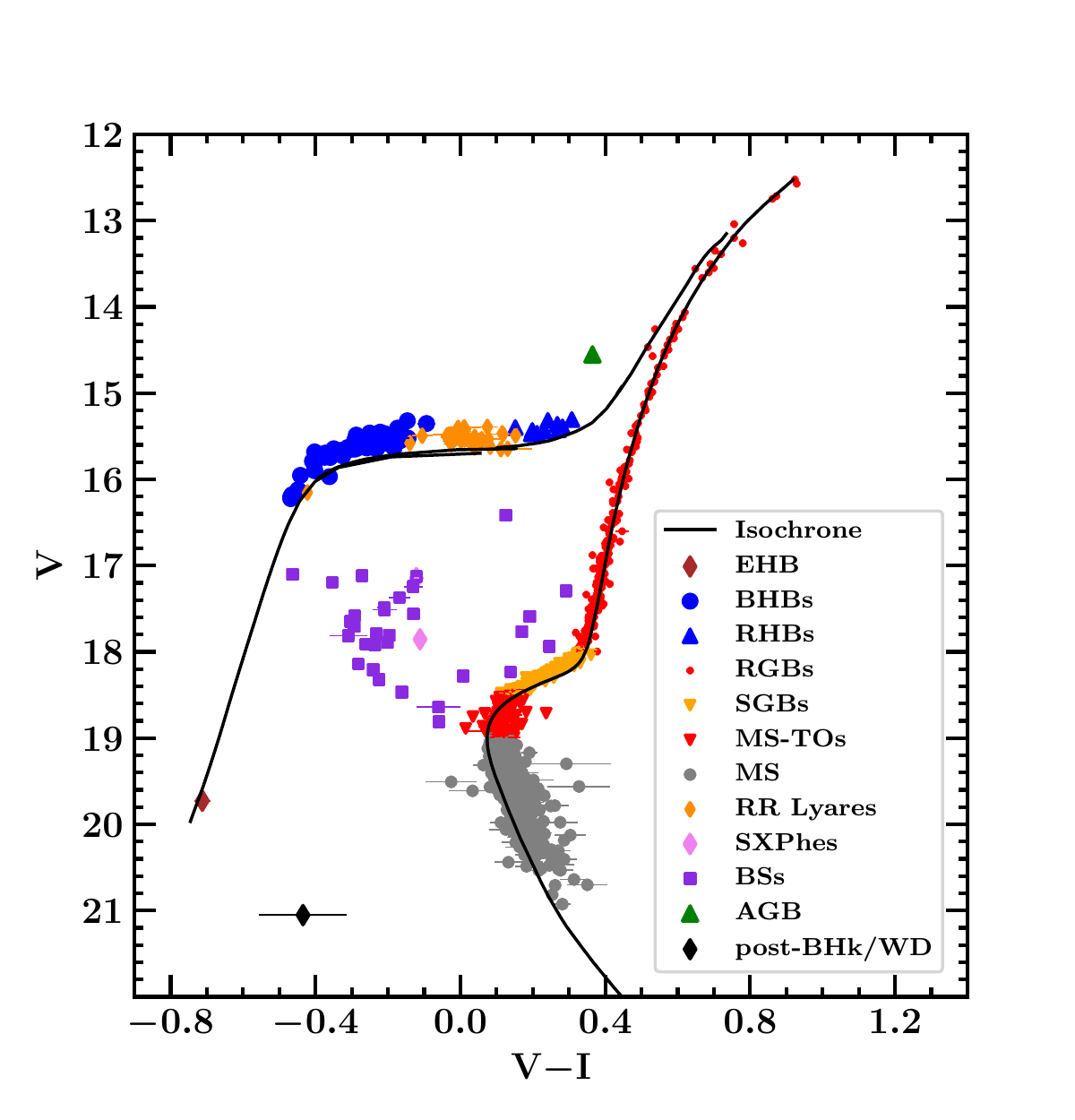}
    \includegraphics[width=0.495\textwidth]{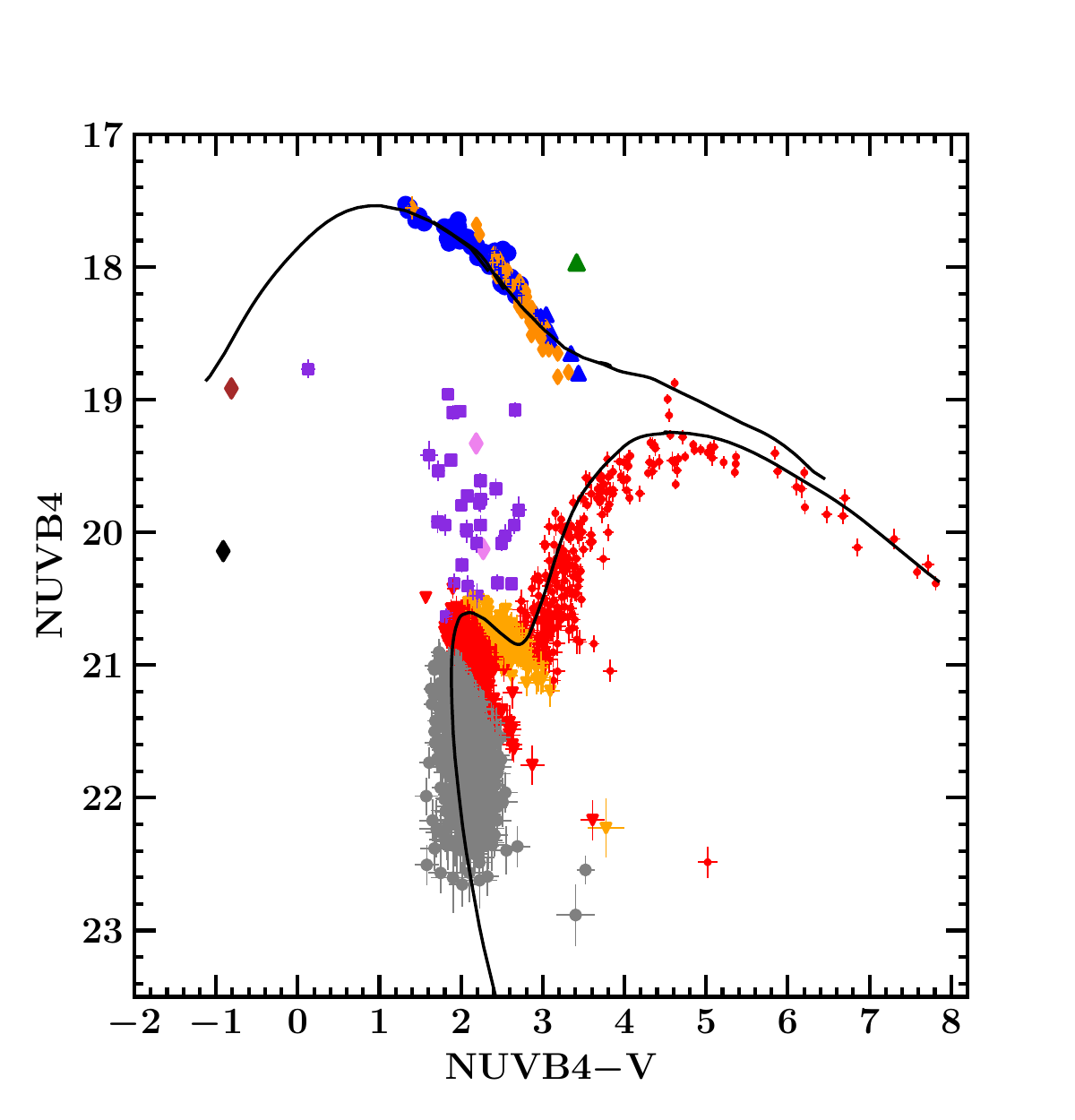}\\
    \includegraphics[width=0.495\textwidth]{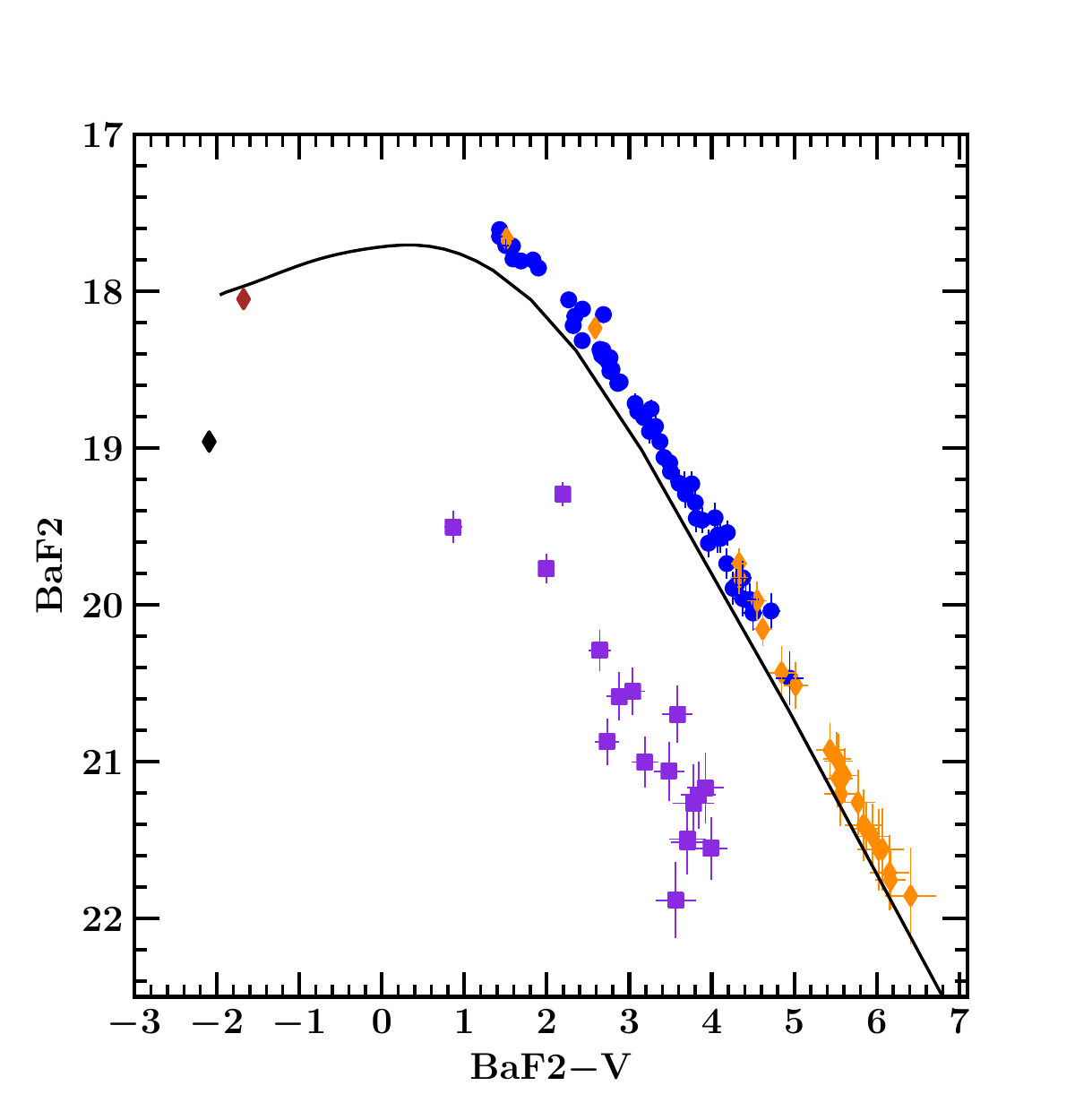}
    \includegraphics[width=0.495\textwidth]{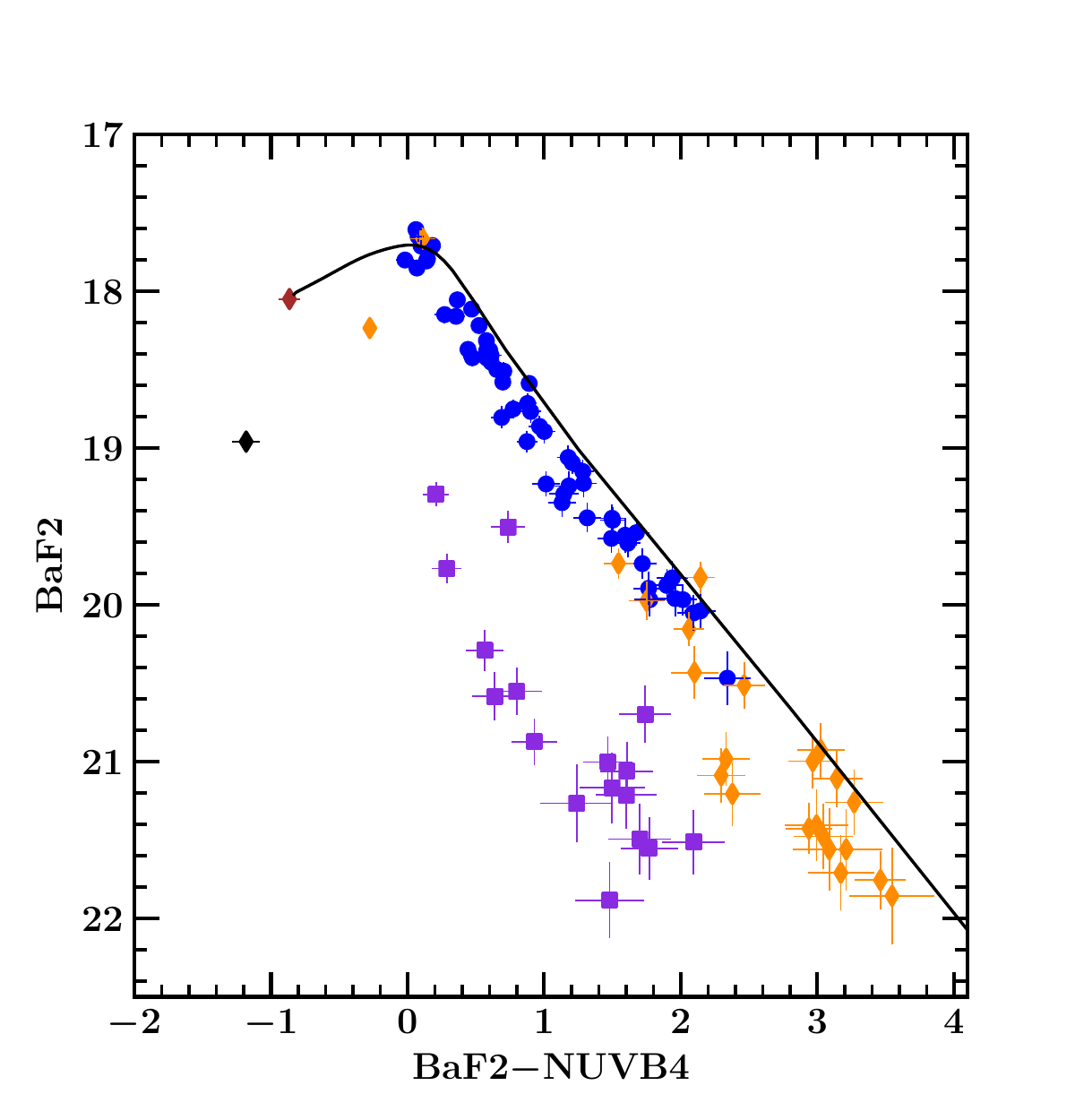}

    \caption{V$-$I vs V (top left), NUVB4$-$V vs NUVB4 (top right), BaF2$-$V vs BaF2 (bottom left), and BaF2$-$NUVB4 vs BaF2 (bottom right) CMDs of the cluster. Sources of different evolutionary stages are denoted with the colors and symbols mentioned in the legend of top left panel.  The BaSTI-IAC model isochrone and ZAHB of chemical composition: $\mathrm{[Fe/H]}=-2.20$, $[\alpha/Fe]=+0.4$ and $Y=0.247$, age = 12.5 Gyr and a distance modulus of $(m-M)_v = 15.21$ are overplayed in black solid line.}
    \label{fig:uv-opt_cmd}
\end{figure*}

We have shown the optical, NUV-optical, FUV-optical, and FUV-NUV CMDs for all the cluster member stars observed with UVIT filters in \autoref{fig:uv-opt_cmd}. We have used the optical and NUV-optical CMDs to identify stars of various evolutionary phases present in the cluster which are marked with various symbols and colors as mentioned in legends of the diagrams. Details of the number of sources of each evolutionary phase are provided in \autoref{tab:evol_phase}. We used an updated BaSTI model\footnote{\href{http://basti-iac.oa-abruzzo.inaf.it/index.html}{http://basti-iac.oa-abruzzo.inaf.it/index.html}} (hereafter: BaSTI-IAC) alpha-enhanced mixture grids \citep{Pietrinferni2020} to generate isochrones and zero-age horizontal branch (ZAHB). The isochrones and the ZAHB models selected for the present analysis correspond to the chemical composition: $\mathrm{[Fe/H]}=-2.20$, $[\alpha/Fe]=+0.4$, $Y=0.247$, and age = 12.5 Gyr. We have over-plotted the isochrones and ZAHB on all the CMDs using solid black line in \autoref{fig:uv-opt_cmd}. We see that the isochrones fit well with the observed CMDs of cluster member stars.

The position of MSTO is well defined in color in optical CMD ($\sim$0.08 mag in $V-I$ color in the top left panel \autoref{fig:uv-opt_cmd}), however it is even more distinct in magnitude in NUV-optical CMDs ($\sim$20.6 mag in NUVB4 mag in the top right panel of \autoref{fig:uv-opt_cmd}). There are about 30 BSs in the region between MSTO and HB. We found one evolved-BS (E-BS) in the core of the cluster. The position of the E-BS is at (0.15, 16.4) mag and (2.8, 19.2) mag in V$-$I vs V and NUVB4-V vs NUVB4 CMDs, respectively (upper panels of \autoref{fig:uv-opt_cmd}). Out of 30 BSs, 25 are detected in FUV filters that we have shown on the FUV-NUV and FUV-optical CMDs (bottom panels of \autoref{fig:uv-opt_cmd}). All the FUV-bright BSs are lying in parallel sequence and around one magnitude below of HBs in FUV-NUV and FUV-optical CMDs. \citet{Simunovic2016} had detected 51 BSs in the core of the cluster using {\em HST} ACS optical filters (F606W and F814W). However, our selection of BSs in the {\em HST} observed region is based on the NUVB4 magnitudes and we considered only those stars which were clearly above the MSTO in the NUVB4-V vs NUVB4 CMD (upper right panel, \autoref{fig:uv-opt_cmd}) and free from the blending effect on the NUV or FUV filter images. This led us to detection of 30 BSs in the {\em HST} observed region\footnote{Here we note that our BSs sample is not complete in the inner region, instead we exploit the UV bright BSs population based upon the UVIT resolved source sample.}.

We found 68 HB stars and one EHB star in the cluster NGC 4590. The BHB stars are visible in both FUV and NUV filters whereas RHB stars are visible only in NUV filters (see \autoref{fig:uv-opt_cmd}). The HB sequence is horizontal in V$-$I vs V CMD consisting of all the HB stars (11 RHB stars and 57 BHB stars), and shows a visible gap filled with the variable stars. We found 41 RR Lyrae stars between BHB and RHB gap by cross-matching the observed HB stars with the variable stars catalog of \citet{Clement2001} within a matching radius of 2$''$.
\cite{Walker1994} has reported 59 BHBs and 11 RHBs in the cluster using optical CMDs. Out of these 59 BHBs, we found that 57 are confirmed non-variable BHBs. For the rest two stars, one was identified as variable star in the variable stars catalog of \cite{Clement2001} and the other one is in the core of the cluster affected by a BSs blend (within 1.5$''$ radius). 

We found two new FUV bright stars in the core of the cluster at a distance 65$''$ from the cluster center (but in different directions) which were not reported previously as cluster members in any optical or {\em HST} studies. However, we find these sources are bright in FUV filters and having membership probability of 97.0\% and 96.5\% in the {\em HST} GCs catalog \citep{Nardiello2018}. From optical and UV-optical CMDs, one source is lying in the EHB region ($\mathrm{RA}=189^\circ.874375$,\ $\mathrm{DEC=-26^\circ.761361}$) and the other one in the WD / bright gap region ($\mathrm{RA}=189^\circ.852542$,\ $\mathrm{DEC=-26^\circ.759667}$) of the cluster. A detailed investigation of these sources will be done in the subsequent sections. The RGB, sub-gaint branch (SGB), and MS stars were selected in the outer part of the cluster using {\em Gaia} proper motions (we excluded the {\em HST} observed region in the core to avoid blending effect while selection). These sources were visible only in NUV filters and gradually being fainter towards lower wavelengths. We also found one star above the RHB in the optical CMD. This source might be either in the post-He core burning phase (pHB) evolving towards AGB phase or in the E-BSs phase \citep{Sills2009}. We detect this star 0.5 mag above the HB sequence in all the NUV filters. The star (AGB) is shown as a green upper triangle in \autoref{fig:uv-opt_cmd}.

\section{Spectral energy distribution (SED) of UV bright stars}
\label{sec:sed}

We have modelled the SED to derive the physical parameters (e.g., effective temperature, bolometric luminosity, surface gravity, etc.) of FUV bright (BHBs, BSs, EHB, and WDs) and NUV bright (RHBs, pHB, and BSs) stars of the cluster. Fitting of SED needs a large spectral coverage of stellar radiation from the UV to the infra-red (IR) wavebands. We have used photometric surveys of UVIT, {\em HST}, {\em Gaia}, {\em GALEX}, Panoramic Survey Telescope and Rapid Response System (Pan-STARRs) survey, 4 m telescope at Cerro Tololo Inter-American Observatory (CTIO-4m, U, B, V, R, I), and Two Micron All Sky Survey (2MASS) for the SED fittings of the observed UV bright stars. The details of the telescopes and their filters in UV, optical and IR bands are given in \autoref{tab:telescope}. We have used UVIT, {\em HST}, and UBVRI photometric fluxes to fit the SEDs of stars in the {\em HST} observed core region of the cluster and used UVIT, {\em GALEX}, UBVRI, {\em Gaia}, Pan-STARRS and 2MASS photometric fluxes to fit the SEDs of stars observed in the outer region ({\em Gaia} confirmed stars).

We used VO SED Analyzer\footnote{\href{http://svo2.cab.inta-csic.es/theory/vosa/}{http://svo2.cab.inta-csic.es/theory/vosa/}} \citep[][hearafter VOSA]{Bayo2008} to produce grids of model fluxes of different stellar atmosphere models (Kurucz stellar atmosphere model, Levenhagen WD model, and Koester WD model) incorporating the response curves of various filters. In order to estimate the best fitted parameters of SED, it uses minimum-$\chi^2$ technique to fit the observed fluxes with the model generated synthetic fluxes. We used the grids of model fluxes from the Kurucz stellar atmosphere model \citep[][hearafter Kurucz model]{Castelli1997} to fit the observed SED fluxes of EHB, BHBs, BSs, and RHBs. 

\begin{figure*}
    \centering
    \includegraphics[width=0.795\textwidth]{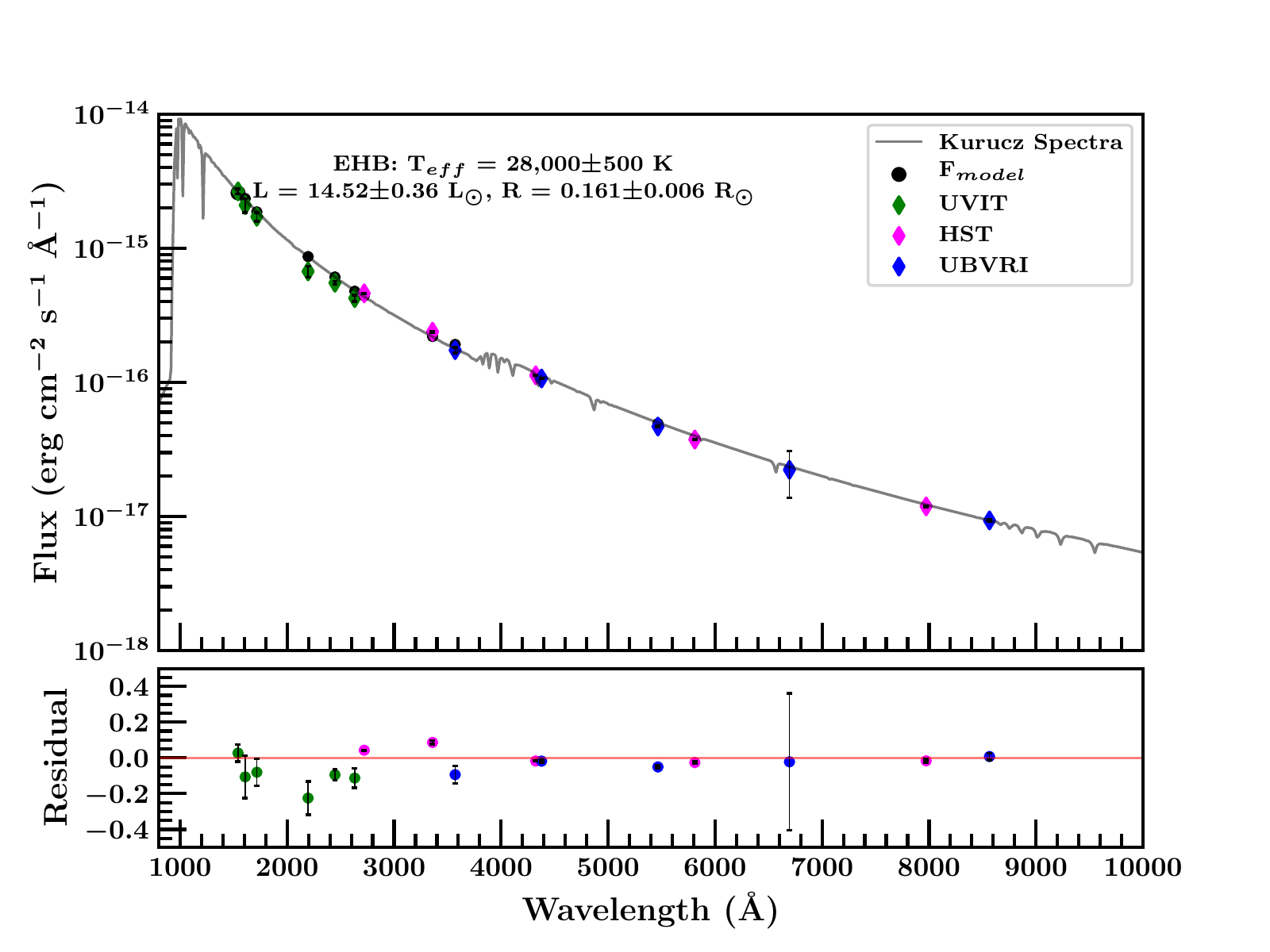}
    \caption{SED of the EHB star. The upper panel shows the observed fluxes (color diamonds) along with the overlaid best fitted Kurucz model spectrum (gray solid line). The black solid circles represent the calculated model fluxes of the respective filters. The lower panel shows the fractional residual between observed and best fitted model fluxes for different filters. The residual is calculated as ($F_{\mathrm{obs.}} - F_{\mathrm{mod}})/F_{\mathrm{mod}}$. The error bar on residuals is calculated as $F_{\mathrm{obs, err}} / F_{\mathrm{obs}}$.}
    \label{fig:sed_ehb}
\end{figure*}

\begin{figure*}
    \centering
    \includegraphics[width=0.495\textwidth]{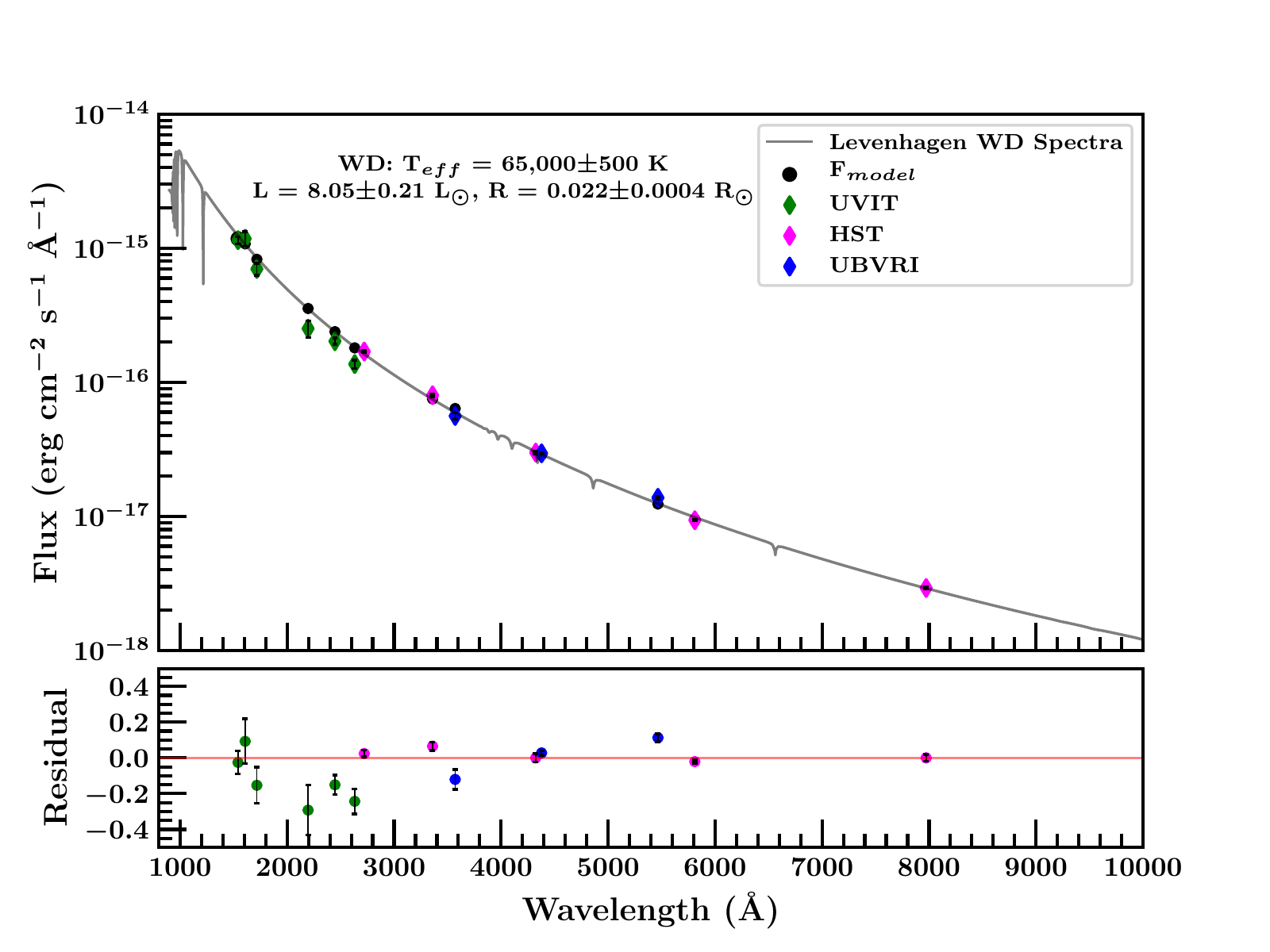}
    \includegraphics[width=0.495\textwidth]{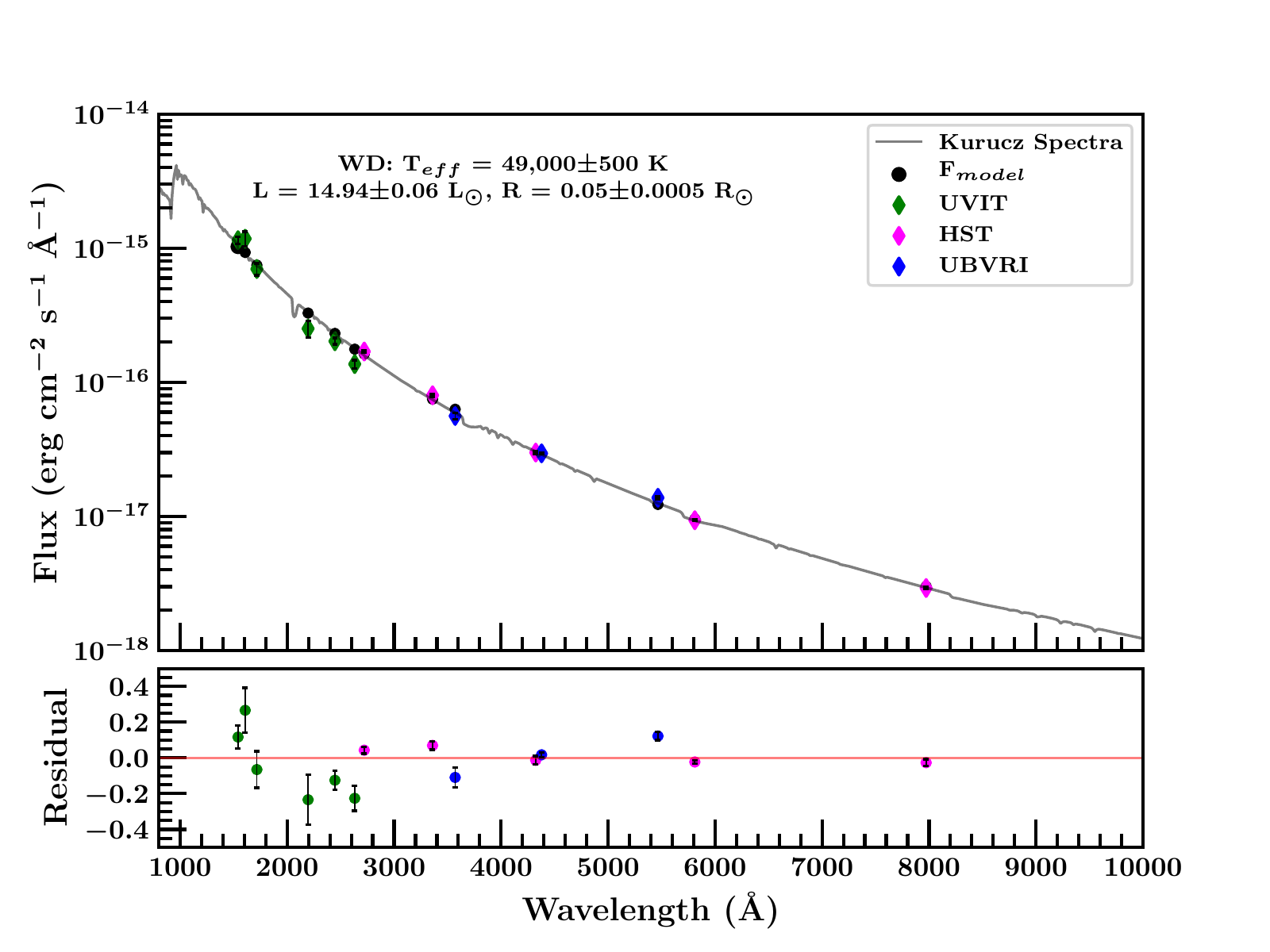}
    \caption{The left panel shows the SED fitting of the newly detected source with Levenhagen WD model whereas the right panel shows the SED fitting of the same source with the Kurucz model. Both the models are fitting well with the observed fluxes. Residuals are shown at the bottom of the each panels.}
    \label{fig:sed_plot}
\end{figure*}

In \autoref{fig:sed_ehb}, we have shown the SED of single EHB star identified in the cluster. The best fitted Kurucz model gives $\mathrm{[Fe/H]}=-2.5$, $\log(g) = 5.0$ and T$_{eff}=28,000$ K. The derived luminosity and radius of the EHB star are 14.52 $\pm$ 0.36 L$_\odot$ and 0.161 $\pm$ 0.006 $R_\odot$, respectively. Similarly, we have used both the Levenhagen WD model \citep[with best fitting parameters, T$_{\mathrm{eff}}=65,000$ K and  $\log(g) = 7.0$, ][]{Levenhagen2017} and the Kurucz model (with best fitting parameters, $\mathrm{[Fe/H]}=0.0$, $\log(g) = 5.0$ and T$_{\mathrm{eff}}=49,000$ K) to the observed spectrum of the newly identified unknown source. The observed flux along with the fitted SED is shown in \autoref{fig:sed_plot}. Undoubtedly, the SED fitting of the source suggests that the star is a very hot source with effective temperature and surface gravity ranges between $50,000 - 65,000$ K and $5.0 - 7.0$, respectively. Hence, we used the mean values of T$_{\mathrm{eff}}$ and $\log(g)$ and errors considering the ranges estimated from the fitted model spectra. We consider a T$_{\mathrm{eff}}$ of 57,500$\pm$7,500 K and $\log(g)$ of 6.0$\pm$1.0 dex to derive luminosity and radius which come out to be  11.50$\pm$3.45 L$_\odot$ and 0.035 $\pm$ 0.015 $R_\odot$ respectively. We have further discussed about its evolutionary status in  \S\ref{sec:evol_status}. 

\begin{figure*}
    \centering
    \includegraphics[width=0.495\textwidth]{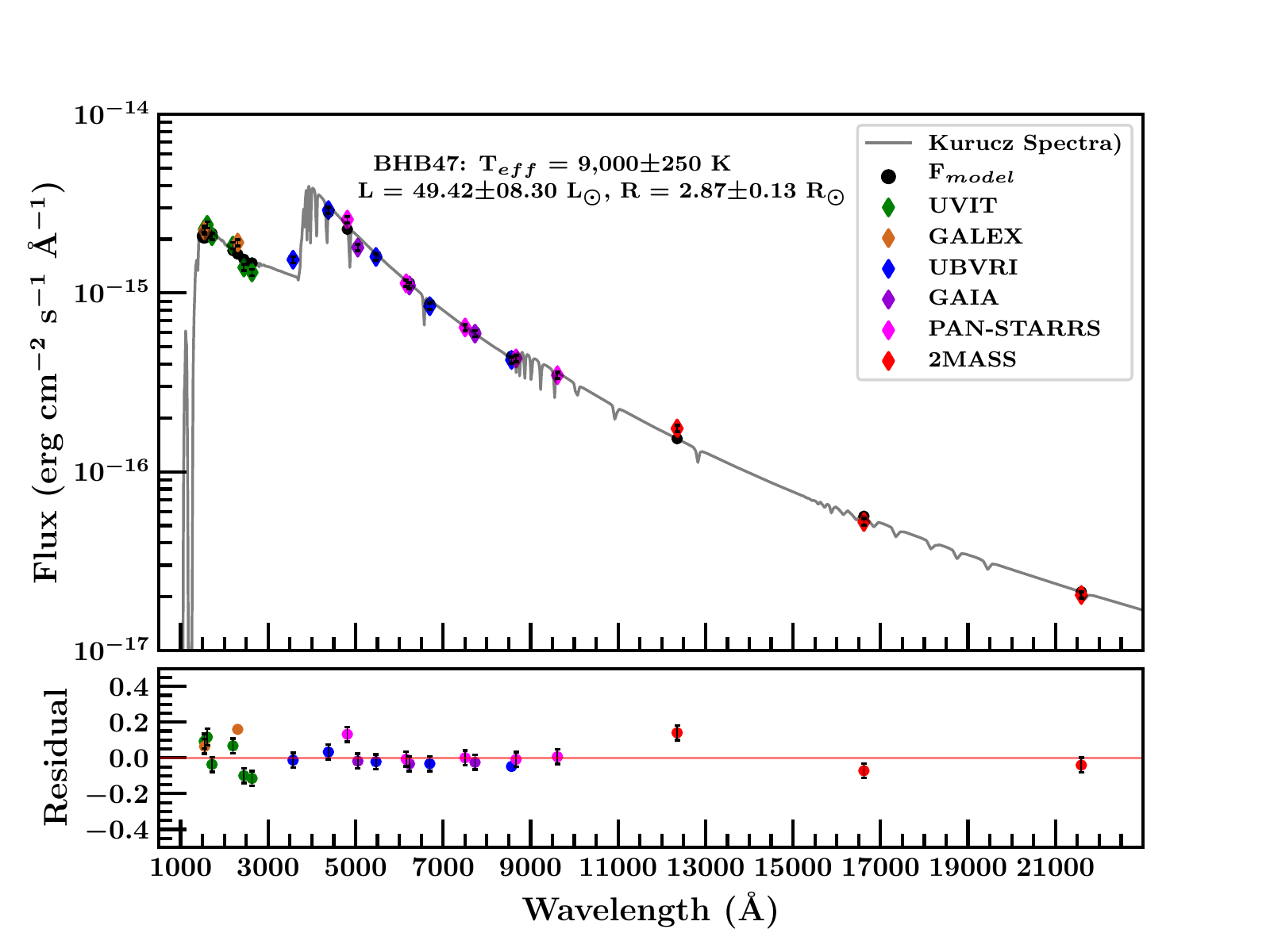}
    \includegraphics[width=0.495\textwidth]{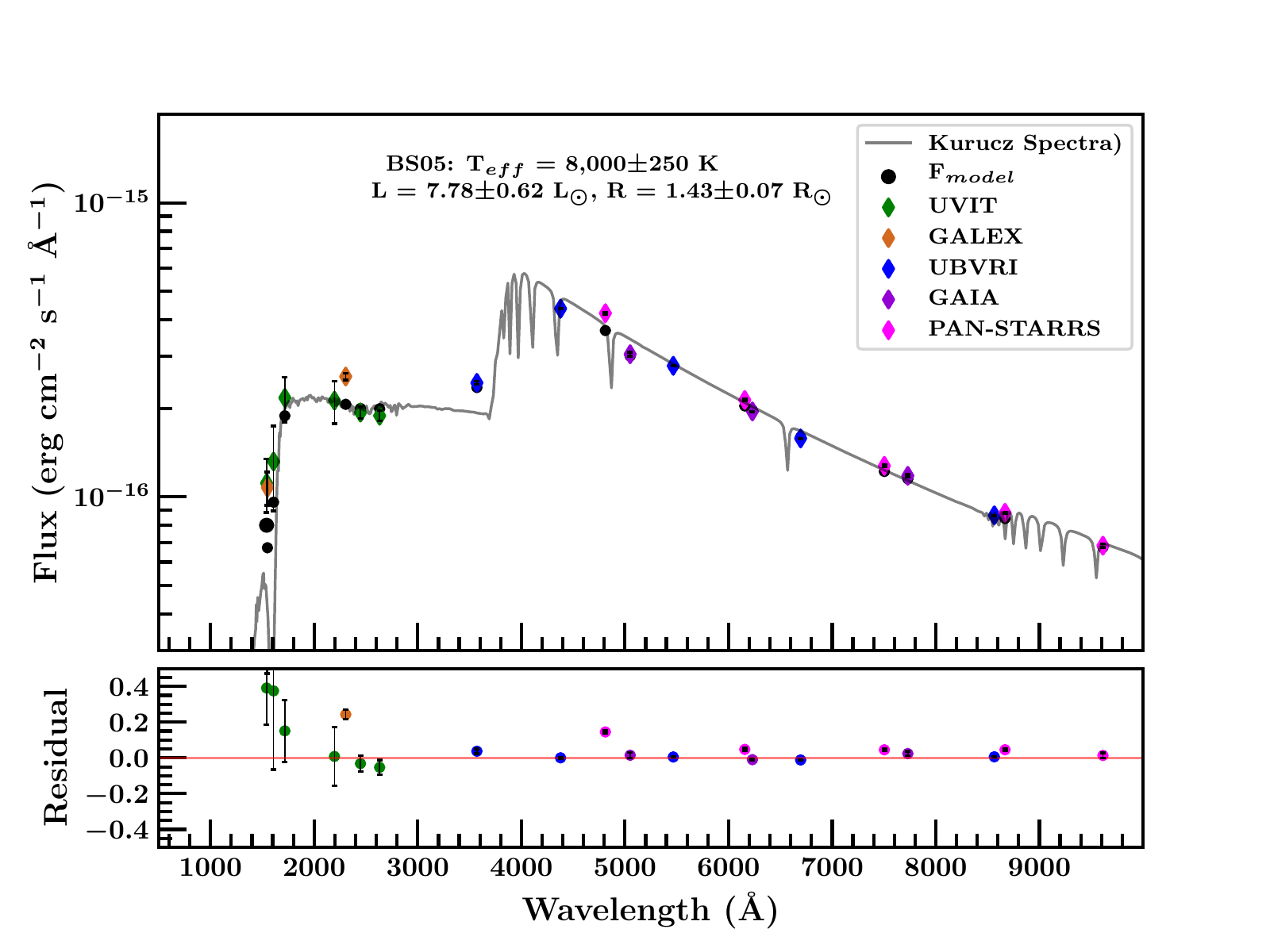}
    \caption{Left panel: SED of the BHB star, right panel: SED of the BS star. The upper panel shows the observed fluxes (color diamonds) along with the overlaid best fitted Kurucz model spectrum (gray solid line). The black solid circles represent the calculated model fluxes of the respective filters. The lower panel shows the fractional residual between observed and best fitted model fluxes for different filters. The residual is calculated as ($F_{\mathrm{obs.}} - F_{\mathrm{mod}})/F_{\mathrm{mod}}$. The error bar on residuals is calculated as $F_{\mathrm{obs, err}} / F_{\mathrm{obs}}$. }
    \label{fig:sed_bhb}
\end{figure*}

\begin{comment}
\begin{figure*}
    \centering
    \includegraphics[width=0.795\textwidth]{SED_BS05.pdf}
    \caption{SED of the BS star (ID: BS05). The upper panel shows the observed fluxes (color diamonds) along with the overlaid best fitted Kurucz model spectrum (gray solid line). The black solid circles represent the calculated model fluxes of the respective filters. The lower panel shows the fractional residual between observed and best fitted model fluxes for different filters. The residual is calculated as ($F_{\mathrm{obs.}} - F_{\mathrm{mod}})/F_{\mathrm{mod}}$. The error bar on residuals are calculated as $F_{\mathrm{obs, err}} / F_{\mathrm{obs}}$ }
    \label{fig:sed_bss}
\end{figure*}
\end{comment}

We have identified 57 BHB stars, of which 33 are in the {\em HST} observed core region and 24 are in the outer region of the cluster.  
We have provided details about all the physical parameters of BHB stars derived from the SED fitting in \autoref{tab:BHB}. Similarly, the physical parameters derived for the RHB stars  and BSs are given in \autoref{tab:RHB} and \autoref{tab:BSs}, respectively. In \autoref{fig:sed_bhb}, we have shown SED of one BHB star (left panel) and one BS star (right panel). We also derived the physical parameters of the NUV filter detected AGB star. Its bolometric luminosity, radius, effective temperature and surface gravity are 176.5$\pm$16.56 L$_\odot$, 14.71$\pm$0.67 R$_\odot$, 5,500 K, and 2.0 dex, respectively.

\begin{figure*}
    \centering
    \includegraphics[width=0.49\textwidth]{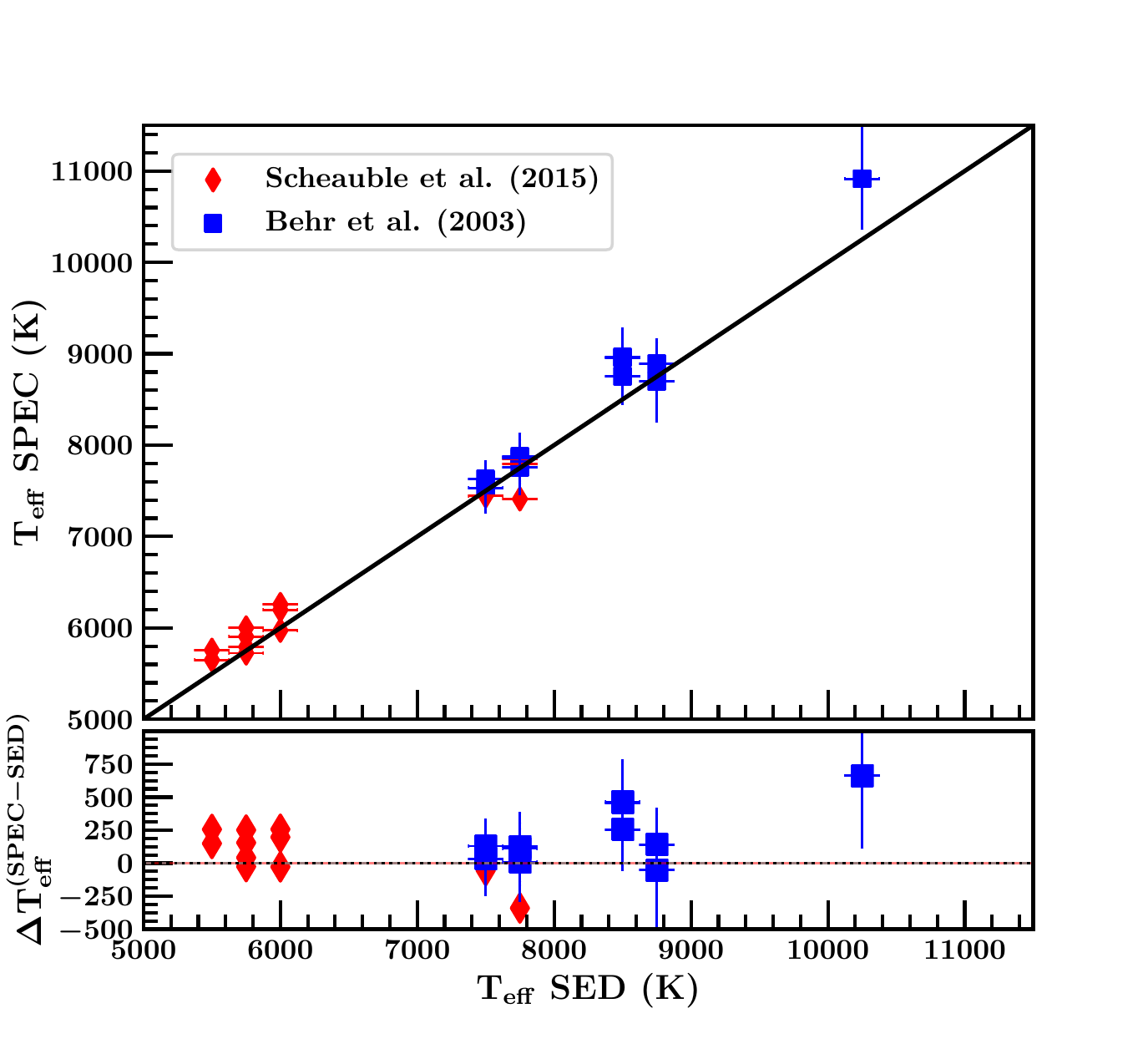}
    \includegraphics[width=0.49\textwidth]{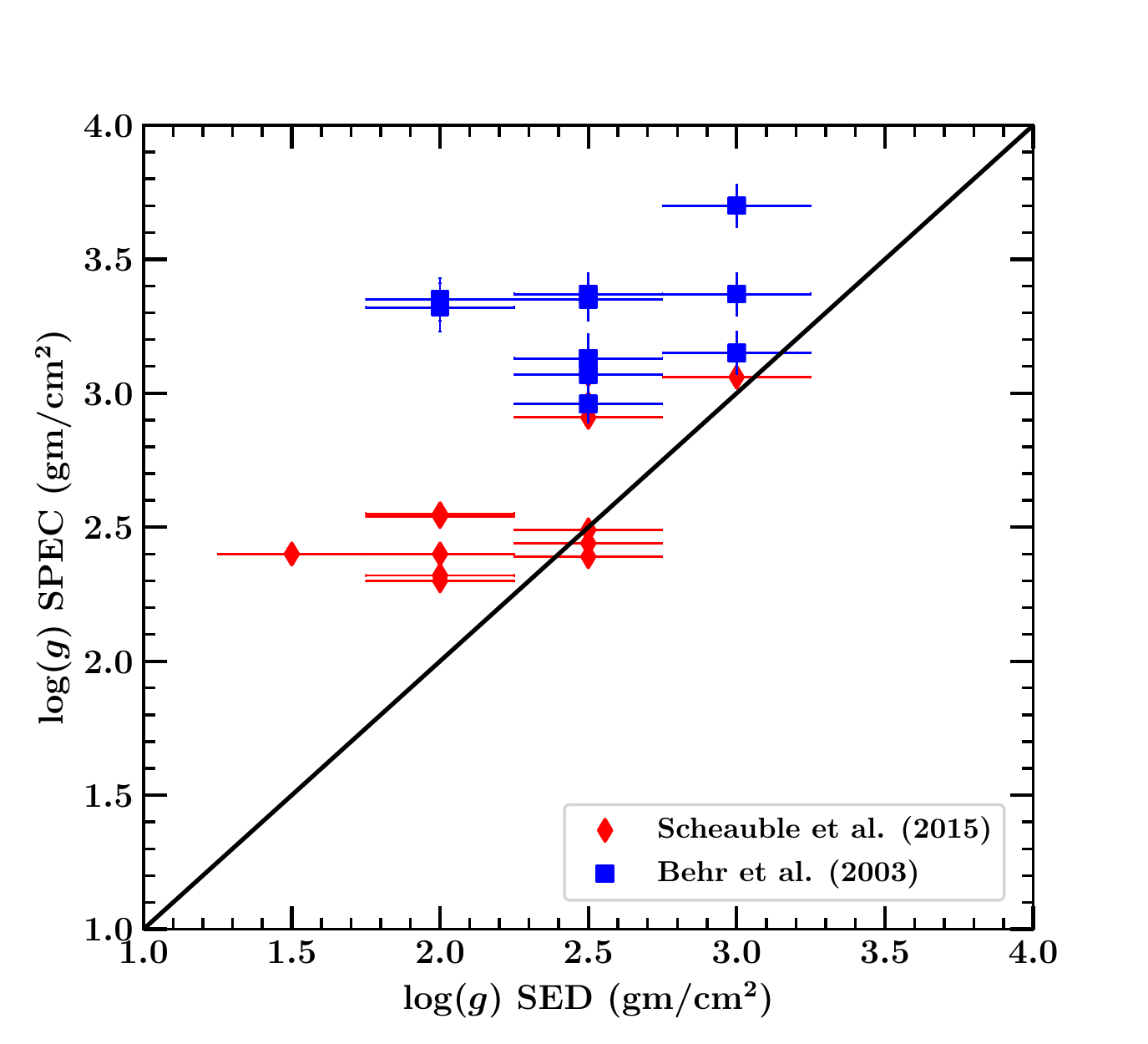}
    \caption{Left panel: comparison of the derived effective temperatures from SED fitting of BHBs and RHBs with the spectroscopic effective temperatures from \citet{Schaeuble2015} and \citet{Behr2003}. Right panel: comparison of the derived surface gravity from SED fitting of BHBs and RHBs with the spectroscopic surface gravity from \citet{Schaeuble2015erratum} and \citet{Behr2003}. }
    \label{fig:teff_comp}
\end{figure*}

The T$_{\mathrm{eff}}$ and $\log$(g) values are spectroscopically determined for five BHBs and nine RHBs by \citet{Behr2003} and for another 11 BHBs by \citet{Schaeuble2015}. We have compared these values with our derived values as shown in \autoref{fig:teff_comp}. We see that the derived T$_{\mathrm{eff}}$ values in this paper are well in agreement with their spectroscopic measured values. However the residuals in the bottom panel of \autoref{fig:teff_comp} shows a difference of 250$-$500 K which is within the errors in SED and spectroscopic estimation. A closer look on the residuals in  T$_{\mathrm{eff}}$ suggests that \citet{Behr2003} derived T$_{\mathrm{eff}}$ is relatively higher than the estimated values from our analysis\footnote{\citep{Behr2003} also noted in their paper that their derived T$_{\mathrm{eff}}$ might be over-estimated and bluer photometry would be needed to constrain the T$_{\mathrm{eff}}$ of hotter BHBs.}. We find that the SED derived $\log$(g) values are relatively lower than the spectroscopic measurements\footnote{Here, we note that \citet{Behr2003} and \citet{Schaeuble2015} derived the $\log$(g) of HB stars only in the optical waveband whereas we have used relatively a larger span of wavebands (UV to IR).}. The HB stars are relatively fainter in the NUV magnitudes (\autoref{fig:uv-opt_cmd}) which might cause a decrease in our derived $\log$(g) values for the HB stars.

\begin{figure*}
    \centering
    \includegraphics[width=0.495\textwidth]{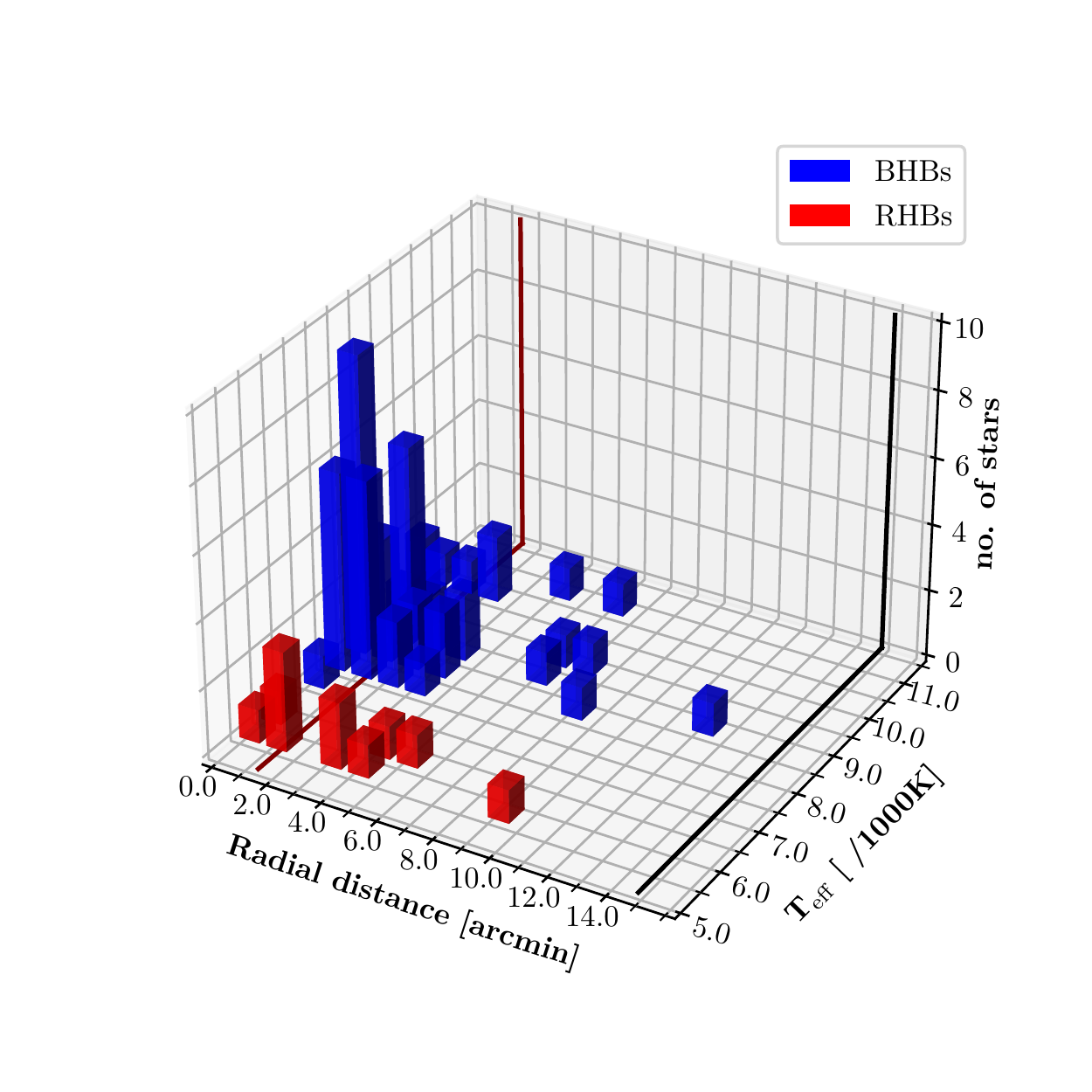}
    \includegraphics[width=0.495\textwidth]{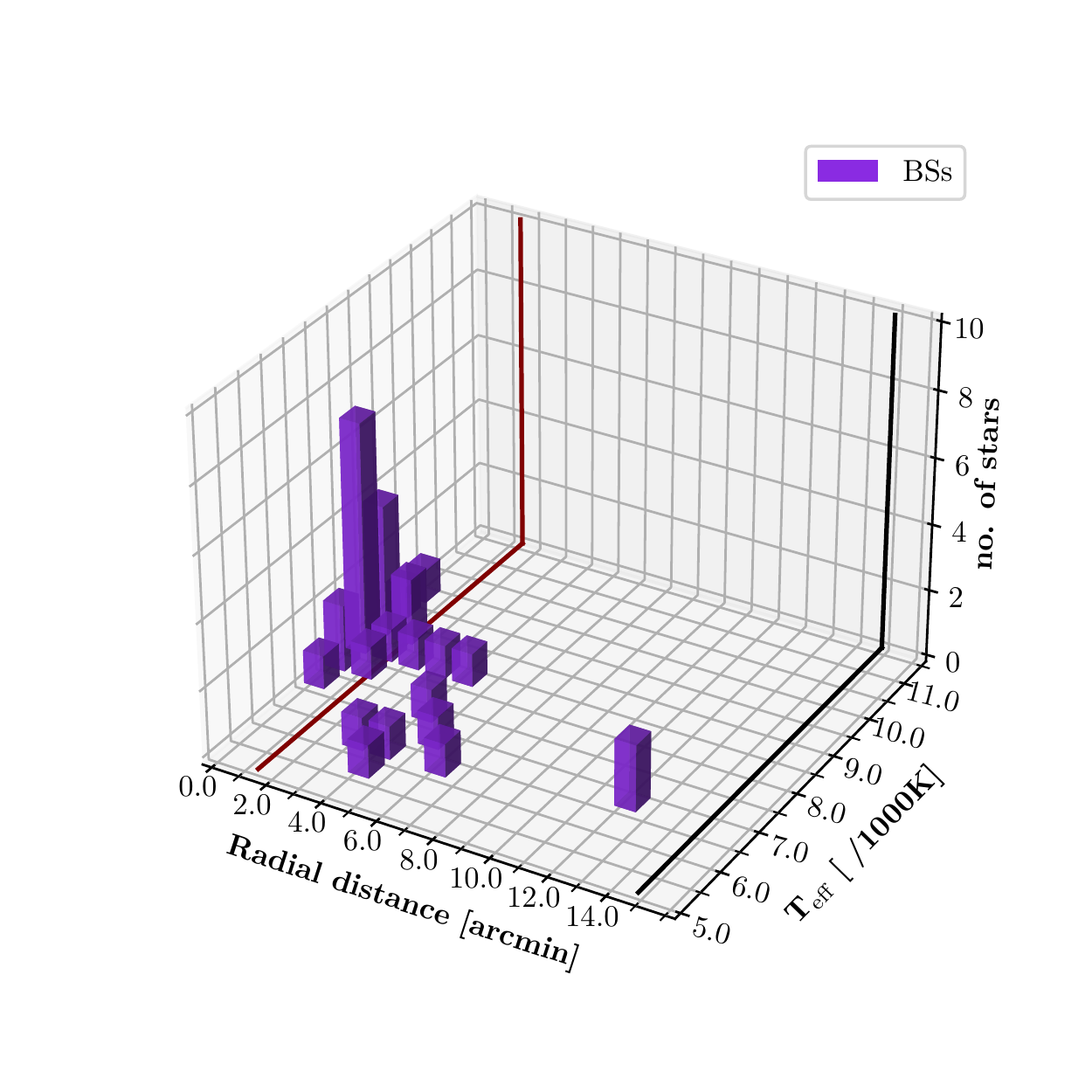}
    \caption{Radial and temperature distribution of HBs (left) and BSs (right). The brown solid line is the half-light radius ($1.51'$) and the black solid line is the tidal radius ($14.91'$) of the cluster.}
    \label{fig:dist_bhb}
\end{figure*}

In the left panel of \autoref{fig:dist_bhb}, we have shown the T$_{\mathrm{eff}}$  and radial distribution of HB stars. The HB stars are mostly distributed in the core (up to 3.5$'$) and a sparse distribution in the outer part of the cluster. All the BHBs detected in this cluster show a temperature range of 7,250 - 10,500 K. The largest population of the BHBs resides within the half-light radius of the cluster (maroon solid line in \autoref{fig:dist_bhb}). However, the RHBs are uniformly distributed up to 9.5$'$ from the cluster center and their T$_{\mathrm{eff}}$ is peaked around 6,000 K (\autoref{tab:RHB}). We also show the radial distribution of BSs in the right panel of \autoref{fig:dist_bhb}. Most of the BSs are situated at the cluster core and spread upto 7$'$. We detected two BSs stars in the outer region of the cluster at a distance of 12$'$ from the cluster center. The T$_{\mathrm{eff}}$ distribution shows a bi-modality which suggests two distinct BSs groups. The redder BSs are situated around 6,000 K whereas the bluer sequence of BSs is seen in the T$_{\mathrm{eff}}$ range of 7,000 K to 9,000 K.

\section{Evolutionary status of UV-bright hot stars}
\label{sec:evol_status}

The UV-bright stars of the cluster are distributed mostly on the HB. The distribution of stars along HB, for a given metalicity and age, is mainly controlled by the mass-loss efficiency during the previous RGB phase and the initial He abundance \citep{Salaris2005}. The temperature of these stars are determined from their SEDs. The stars with T$_{\mathrm{eff}} \sim 6,000$ K are distributed on the RHB whereas stars with T$_{\mathrm{eff}}\ 7,250-10,500$ K are distributed on the BHB, and one star with T$_{\mathrm{eff}}$ 28,000 K is lying on the EHB region. We also find a hot star with T$_{\mathrm{eff}}$ 57,500$\pm$7,500 K which might be an evolved star from the HB phase.

\subsection{Evolution of HB stars}
\begin{figure}
    \centering
    \includegraphics[width=\columnwidth]{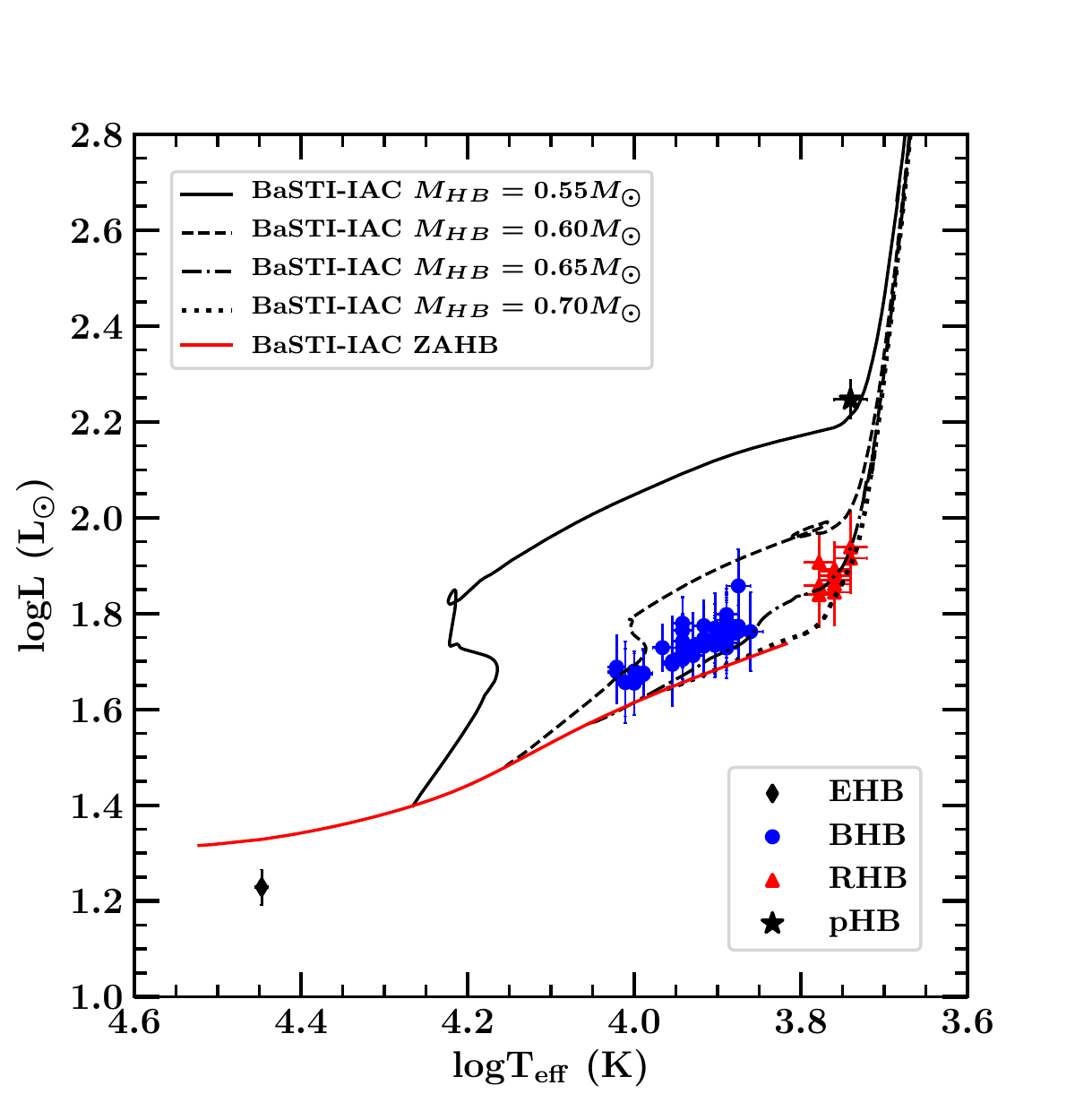}
    \caption{Luminosity vs. effective temperature for EHB, BHB, RHB and post-HB (AGB) stars. BaSTI-IAC ZAHB locus(red solid line) and post-HB evolutionary tracks for 0.55 M$_\odot$, 0.60 M$_\odot$, 0.65 M$_\odot$, and 0.70 M$_\odot$ are over-plotted in black solid, dashed, dash-dotted, and dotted lines, respectively. Colors and shapes of the symbols represent the sources as mentioned in the legend.} 
   \label{fig:lum_teff_hb}
\end{figure}

In \autoref{fig:lum_teff_hb}, we have plotted the physical parameters (L$_{\mathrm{bol}}$ and T$_{\mathrm{eff}}$) derived from the SED fit (color solid points) and the theoretical BHB evolutionary tracks (lines) from BaSTI-IAC models \citep{Pietrinferni2020} for stars of different ZAHB masses. The observed BHBs show a location close to the BaSTI-IAC ZAHB locus corresponding to an RGB progenitor with mass equal to $0.8M_\odot$ (see \citet{Pietrinferni2020} for more details) (red solid line) whereas the unique EHB star is below the ZAHB locus. On the other hand, RHB stars appear to be cooler than the coolest point of the theoretical ZAHB locus. Since, for a given metallicity, the larger is the mass of the HB stars, the cooler is its ZAHB location, the latter evidence could be easily accounted for by assuming that the observed RHB stars or at least a fraction of them are the progeny of BSs. In fact, since the mass of BSs is larger that the typical mass of the stars evolving in a GC ($\sim0.8M_\odot$), this could easily explain the fact that their progeny during the core-He burning stage have a ZAHB effective temperature lower than that expected for the coolest ZAHB point for an old stellar population, corresponding to a mass of about $0.8M_\odot$ (see the discussion in \citet{Cassisi2013} and references therein).

The post-HB evolutionary tracks shown in the same figure, suggest that the HB stars with M$_\mathrm{{ZAHB}}=$ 0.55 M$_\odot$ and 0.60 M$_\odot$ (black solid and dashed lines) during their post-HB evolution cross the H-R diagram at relatively higher luminosity than the evolutionary tracks corresponding to M$_\mathrm{{ZAHB}}=$ 0.65 M$_\odot$, and 0.70 M$_\odot$ (black dash-dotted and dotted lines). The location of the evolutionary tracks near the observed distribution of BHB and RHB stars suggest that the observed BHBs should span a mass range between M$_\mathrm{{ZAHB}}=$ 0.60 and 0.65 M$_\odot$, while and the RHB stars mass range is between M$_\mathrm{{ZAHB}}=$ 0.65 and 0.70 M$_\odot$. The physical parameters derived for the post-HB (AGB) star (black asterisk) are also shown: its position in the H-R diagram is well matched by the post-HB evolutionary phase of a M$_\mathrm{{ZAHB}}\sim$ 0.55 M$_\odot$ stellar model.

\subsection{Evolution of newly detected hot star}

\begin{figure*}
    \centering
    \includegraphics[width=0.54\textwidth]{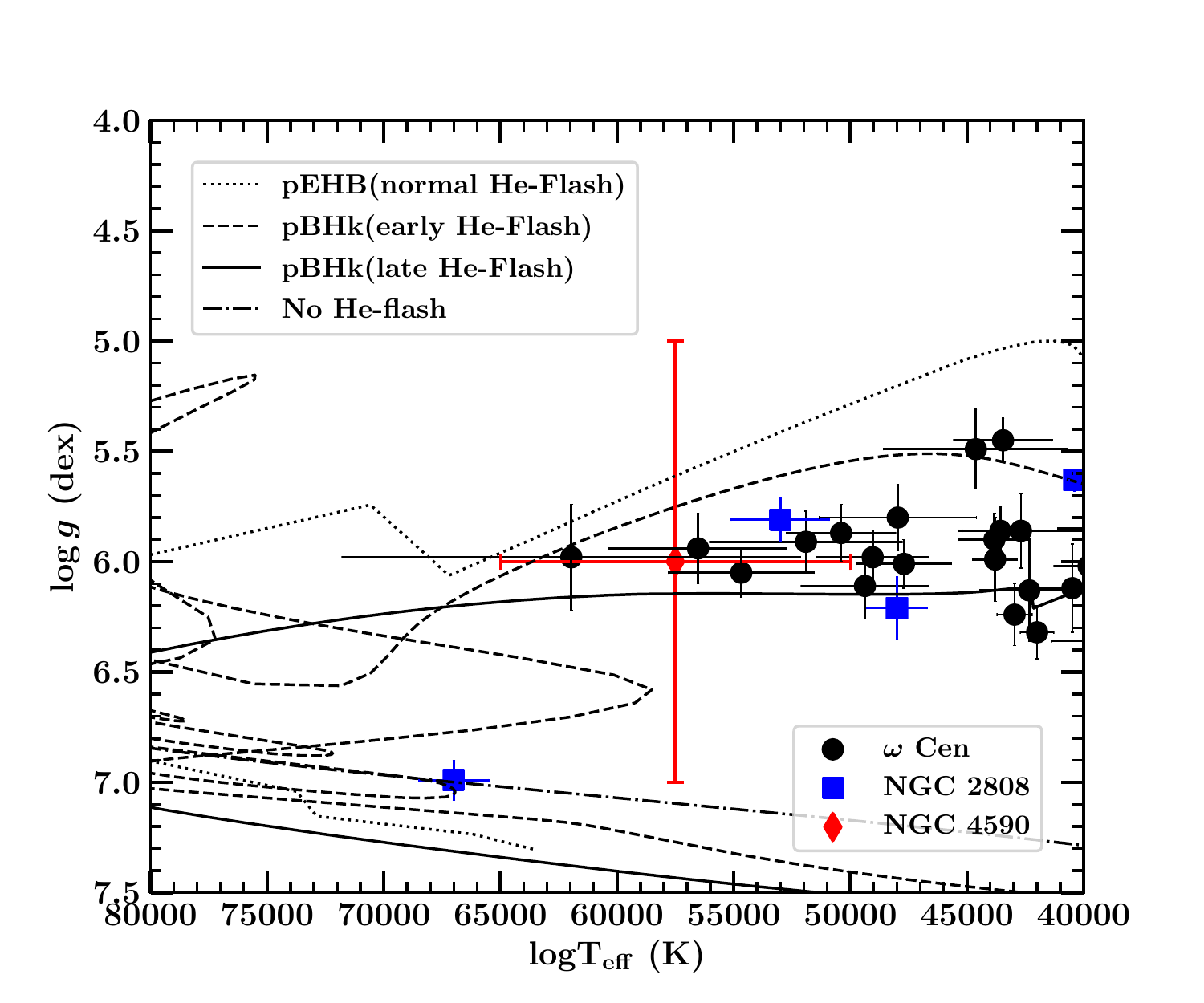}
    \includegraphics[width=0.445\textwidth]{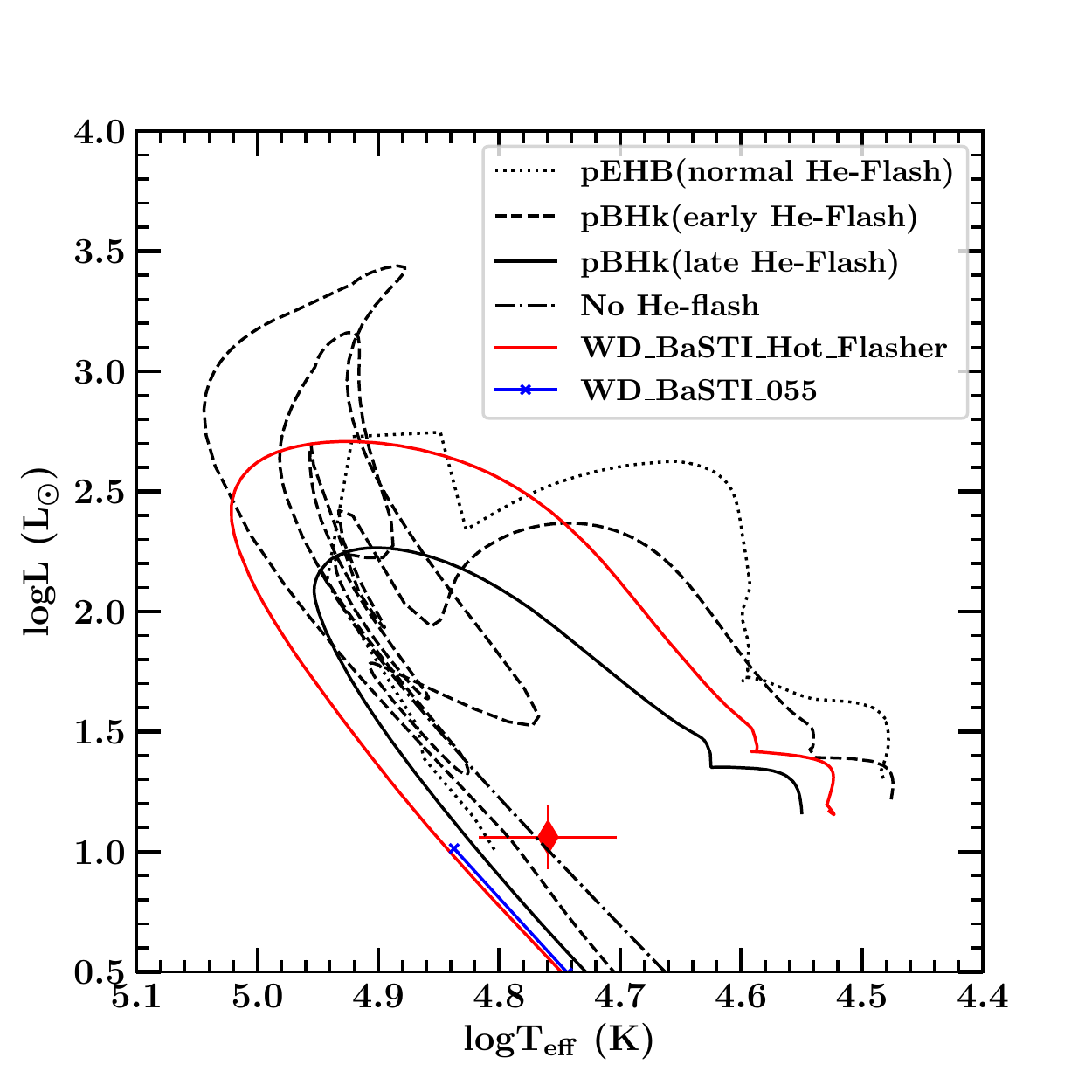}    
    \caption{Left panel: T$_{\mathrm{eff}}$ vs. $\log$(g) of post-BHk stars. The lines show the  evolutionary tracks as described in the legend. The solid circles, squares and diamonds represent the observed values of post-BHk stars in $\omega$ cen, NGC 2808, and NGC 4590, respectively.  Right panel: T$_{\mathrm{eff}}$ vs $\log$L for the newly detected post-BHk or probable WD star. We have also plotted the Hot-flasher post-He core burning evolutionary tracks along with the 0.55 M$_\odot$ WD track to identify the evolutionary status of the newly detected probable WD.}
    \label{fig:lum_teff}
\end{figure*}

The derived T$_{\mathrm{eff}}$ and luminosity of the newly detected hot star support their O-type sub-dwarf (sdO) nature \citep{Heber2016}. Such hot sdO stars are rare in GGCs as well as in the Galactic fields. So far, they have been detected only in two dense GGCs, NGC 2808 \citep{Moehler2004, Brown2012, Brown2013} and NGC 5139 ($\omega$ Cen) \citep{Latour2017, Brown2013}. \citet{Latour2018} suggested that these stars would be in the post-BHk evolutionary phases based on the post-HB evolutionary tracks of hot-flasher models \citep{Bertolami2008} on the T$_{\mathrm{eff}}$ vs $\log$(g) plane. The hot flasher models assume a very efficient mass loss process during the RGB that prevents the He-flash at the tip of RGB and the He-flash occurs either at a higher T$_{\mathrm{eff}}$ while approaching the WD cooling sequence (the early hot flasher scenario) or while cooling along the WD sequence (late He flasher scenario) \citep{Brown2001, Cassisi2003, Bertolami2008}. The very efficient mass loss was proposed to be due to the evolution in a binary system as a consequence of an event of Roche Lobe overflow. \citet{Lei2015} and \citet{Lei2016} studied the hot flasher scenario using binary evolution at the RGB phase with different orbital periods, mass-ratio of binary stars, metallicity and He-abundance variation of the primary star. They found that for stars with Z=0.0003, M$_{\mathrm{ZAMS}} =0.81 M_\odot$, Y=0.24, (similar to the NGC 4590 parameters) would need orbital period of 1560 - 1700 days for early flasher scenario, 1290 - 1560 days for late flasher, and with orbital period less than 1290 days will never go through He-flash and end up as a WD star \citep{Lei2016}. 

In \autoref{fig:lum_teff}, we have plotted post-BHk evolutionary tracks from \cite{Lei2015} for the above discussed the hot flasher scenario. We have also shown the canonical post-EHB evolutionary track (dotted line) of $\mathrm{M_{HB}}=$ 0.505 M$_\odot$ from \citet{Moehler2019}. In the left panel, we have plotted the sdO stars (T$_{\mathrm{eff}}$ > 40,000 K) of NGC 2808 and $\omega$ Cen along with the newly detected hot star of NGC 4590 on the T$_{\mathrm{eff}}$ vs $\log$(g) plane. The T$_{\mathrm{eff}}$ and $\log$(g) for NGC 2808 and $\omega$ Cen were derived using optical spectroscopy by \citet{Moehler2004} and \citet{Latour2018}, respectively. We see that the stars around 50,000 K and above are lying close to the $\log$(g)=6.0 dex. The derived average values of the T$_{\mathrm{eff}}$ and $\log$(g) suggest that the newly detected hot star is lying in between the early and late post-BHk tracks. The other sdO stars of NGC 2808 and $\omega$ Cen are also lying in-between the early and late hot flasher evolutionary tracks. We find a hot star with T$_{\mathrm{eff}}\ \sim 67,000$ K and $\log$(g)=7.0 dex in NGC 2808 and shown in \autoref{fig:lum_teff}. \citet{Moehler2004} suggested that the star might be a low-mass WD appearing at a lower distance than the NGC 2808. Our best SED fitted model (left panel in \autoref{fig:sed_plot}) also suggests that the newly detected hot star, having similar T$_{\mathrm{eff}}$ and $\log$(g), could be a WD. 

The right panel of \autoref{fig:lum_teff} shows that the observed hot star (red diamond) is lying close to the end of post-HB tracks (dotted lines) and beginning of the WD track (blue solid line). This suggests that the observed source is evolving towards WD phase and nearing to complete its post-HB evolutionary phase. Since the source is lying almost on the `no helium-flash' evolutionary track (dash-dotted line) which suggests possibility of no He-flash scenario for the source. However, a large error range in the T$_{\mathrm{eff}}$ also covers the other proposed hot flasher scenario. 

The above analysis suggests that the newly detected hot star in NGC 4590 might be at such phase where it is ending its post-BHk evolutionary phase and about to start its WD cooling phase. However, a high resolution UV spectroscopy is needed to confirm its present evolutionary status. 
\subsection{Evolution of BSs}

There are two commonly accepted channels for the BSs formation: mass-transfer in the binary system \citep{McCrea1964} and a product of direct collision of two or more stars in the contact \citep{Hills1976}. The mass-transfer mechanism dominates in the low density environment, whereas the collisional mechanism dominates in the dense environment. The evolution of BSs is similar to the single star evolution. However, depending upon the environment during merger or collision, the mass and age of BSs slightly differ from the single star evolution \citep{Sills2009, Sun2021}. 

\begin{figure*}
    \centering
    \includegraphics[width=0.54\textwidth, height=0.35\textheight]{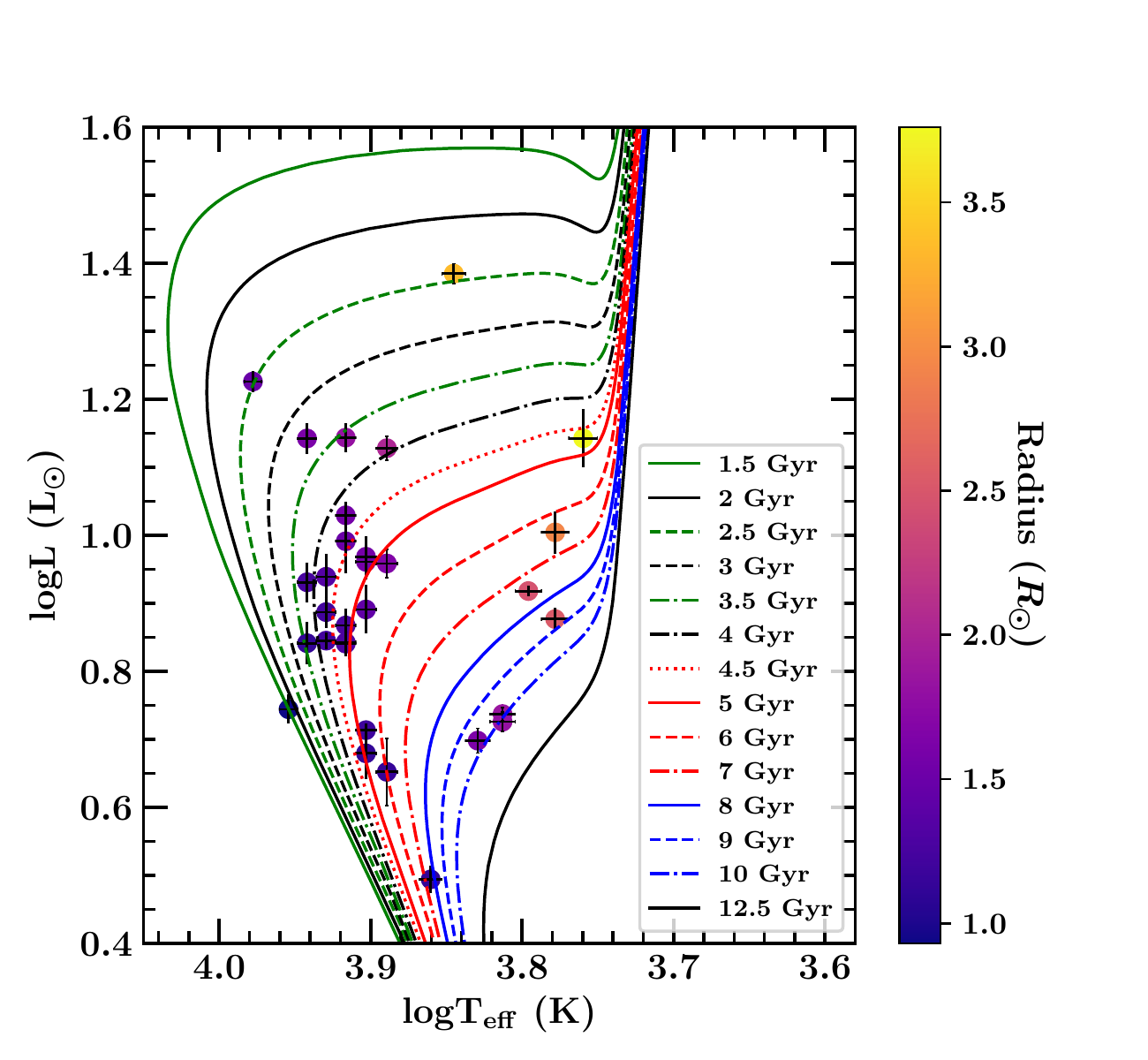}
    \includegraphics[width=0.455\textwidth]{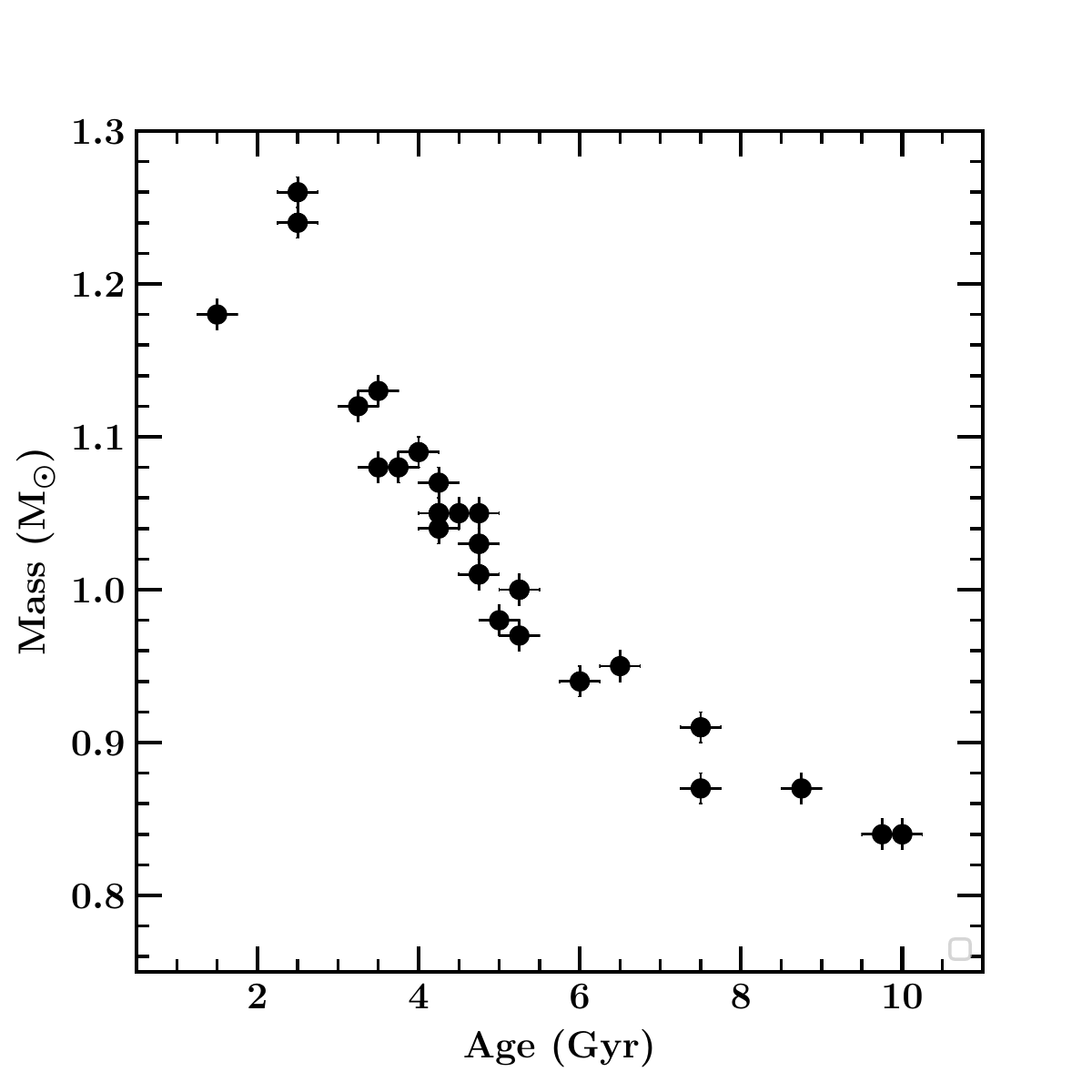}
    \includegraphics[width=0.495\textwidth]{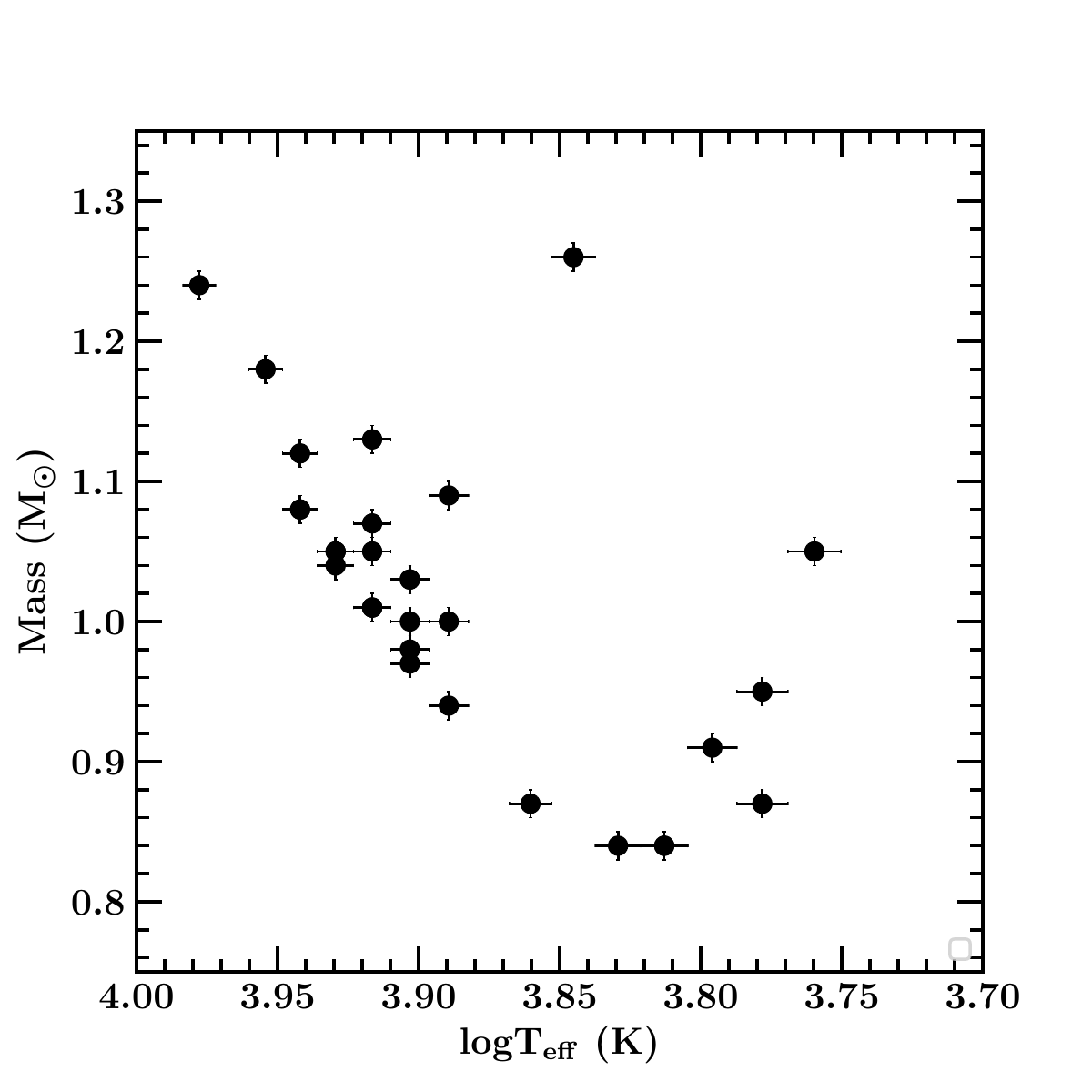}
    \includegraphics[width=0.495\textwidth]{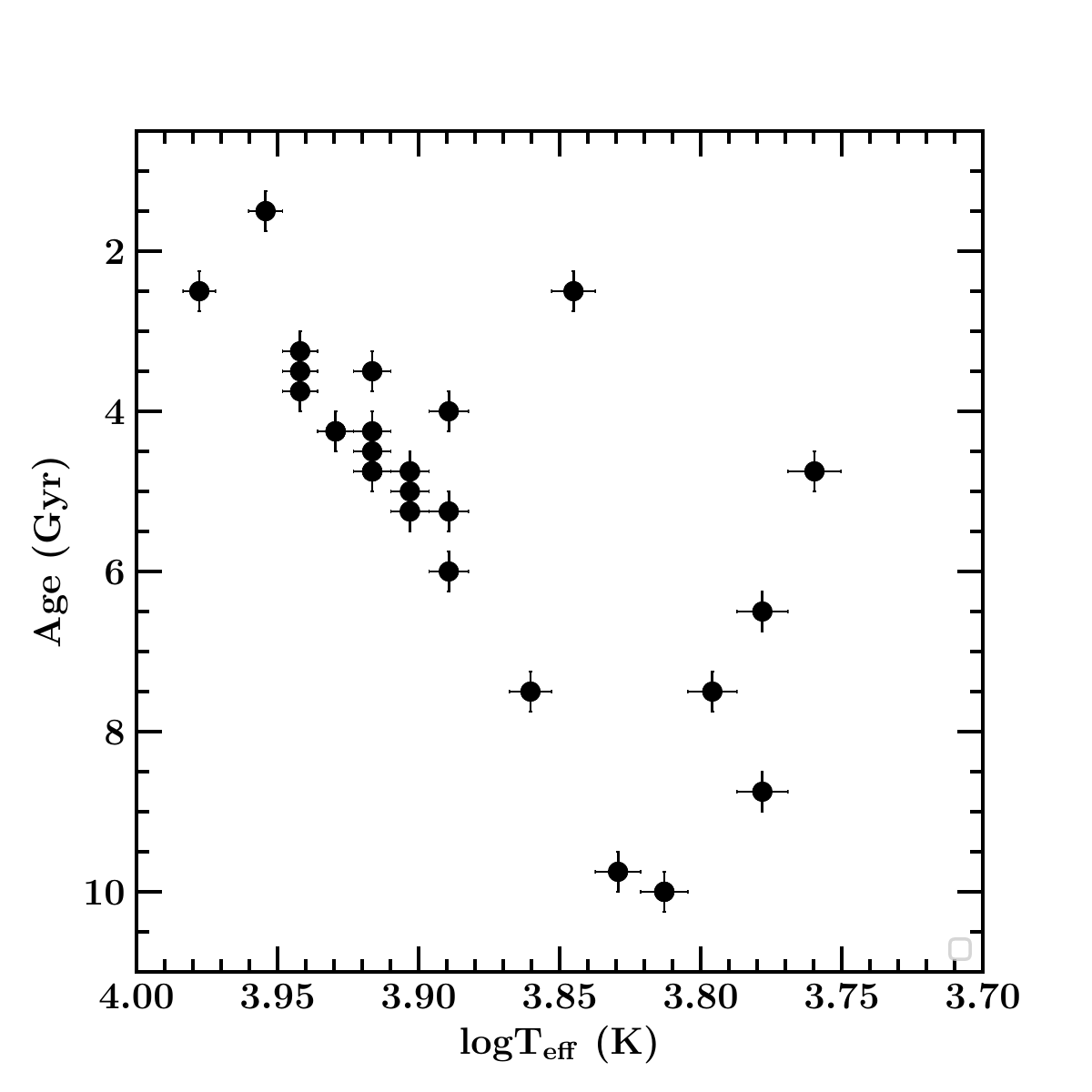}
    
    \caption{Top left panel: the luminosity vs. effective temperature plot for all the 31 BSs. The lines in different colors are BaSTI-IAC isochrones of different ages as shown in the legend. The color bar indicates the radii of BSs derived from the SED fittings. Top right panel: age vs. mass of BSs derived from the isochrone fitting on the T$_{\mathrm{eff}}$ vs. L$_{\mathrm{bol}}$ plane. The bottom left and right panels show the distribution of observed BSs on the T$_{\mathrm{eff}}$ vs. mass and the T$_{\mathrm{eff}}$ vs. age plane, respectively.}
    \label{fig:teff_bs}
\end{figure*}

In the top left panel of \autoref{fig:teff_bs}, we have shown the T$_{\mathrm{eff}}$ and L$_{\mathrm{bol}}$ of the 31 BSs derived from the SED fitting. BaSTI-IAC isochrones spanning a range of ages have also been plotted to study their evolutionary status. We see that most of the BSs are of the age between 3-6 Gyrs and lying on the MSTO region of the isochrones. The red BSs are lying on the 6-10 Gyr isochrones and seem evolved from the MS towards the SGB phase. We find the E-BS star is lying on the 2.5 Gyr age isochrone and seems to be in its SGB phase. We have also plotted 12.5 Gyr isochrone similar to the cluster age for comparison. We see that all the detected BSs are lying above the 12.5 Gyr isochrone. We notice the stellar radii of BSs are ranging in between 1 and 3.5 R$_\odot$ with hotter BSs being smaller and cooler BSs having larger radii. 

Based upon the position of the observed BSs near the relevant BaSTI-IAC isochrones, we derived their mass and age which are listed in \autoref{tab:BSs}. Considering the errors in the T$_{\mathrm{eff}}$ and L$_{\mathrm{bol}}$, we find an uncertainty of 0.01 M$_{\odot}$ and 0.25 Gyr in the derived mass and age of BSs. In the top right panel of \autoref{fig:teff_bs}, we have plotted the derived age and mass of all the 31 BSs. The bottom panels of \autoref{fig:teff_bs} show the distribution of BSs in T$_{\mathrm{eff}}$ vs stellar mass (left bottom panel) and T$_{\mathrm{eff}}$ vs age planes (right bottom panel). 

\begin{figure*}
    \centering
    \includegraphics[width=0.495\textwidth]{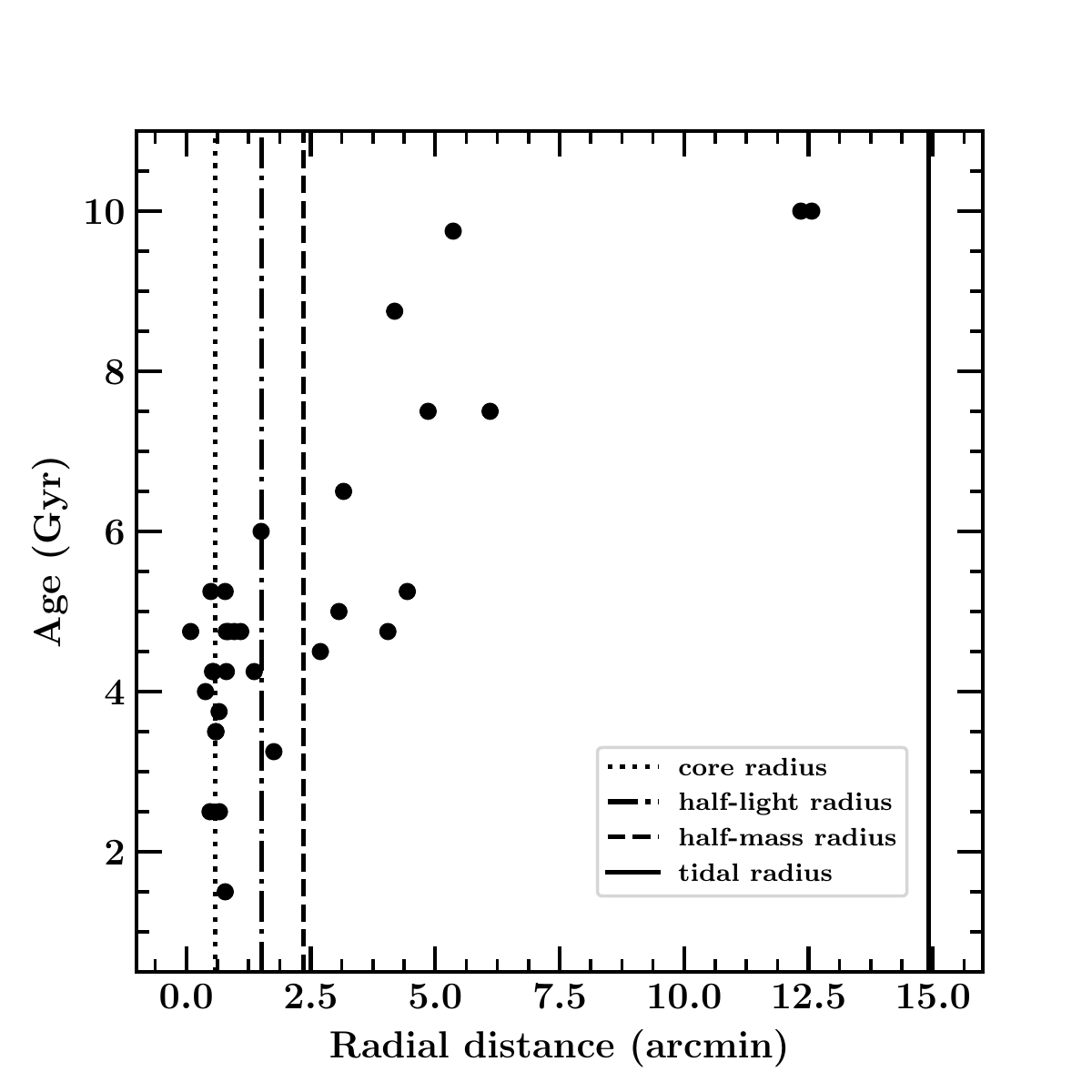}
    \includegraphics[width=0.495\textwidth]{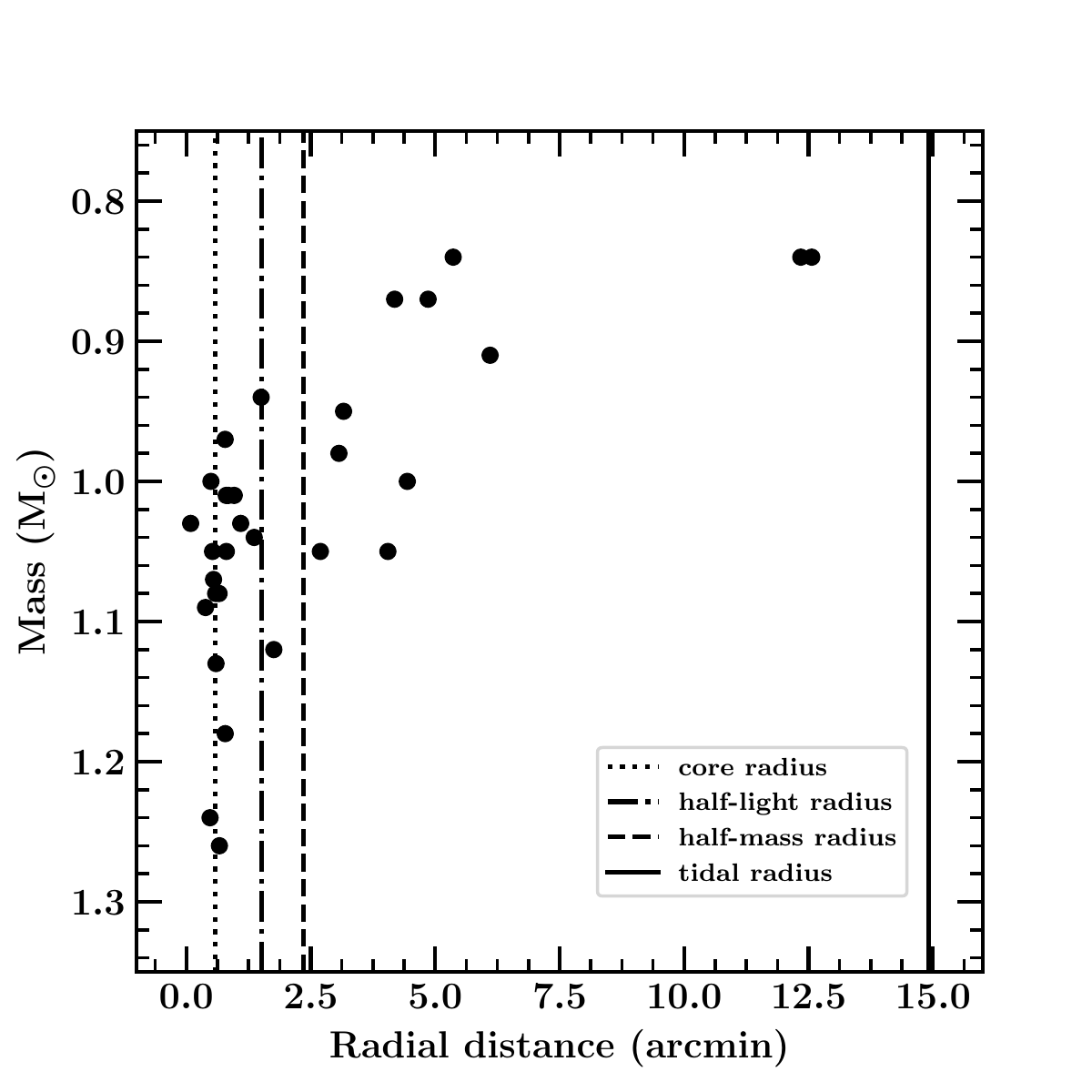}
    \caption{Radial distribution vs. age (left panel) and radial distribution vs mass (right panel) for all the 31 BSs. The vertical lines are plotted for the core radius (dotted line at 0.58$'$), the half-light radius (dash-dotted line at 1.51$'$), the half-mass radius (dashed line at 2.358$'$), and the tidal radius (solid line at 14.91$'$) of the cluster.}
    \label{fig:radial_bs}
\end{figure*}

In \autoref{fig:radial_bs}, we have shown the radial distribution versus age (left panel) and mass (right panel) plots for all the observed 31 BSs. We find that younger and massive BSs are concentrated at the core of cluster whereas older and less massive BSs are distributed up to 14 arcmin from the cluster center. 

All the above presented results suggest that both the mass-transfer as well as collision/merger mechanisms are active in the cluster and massive BSs are formed through collision/merger of massive stars at the center of the cluster whereas the less massive and older BSs are formed through mass-transfer from their companion stars in the binary system.  

\section{Radial distribution of Blue Straggler stars and dynamical age of the cluster}
\label{sec:rad_BSs}
\begin{figure*}
    \centering
    \includegraphics[width=0.49\textwidth]{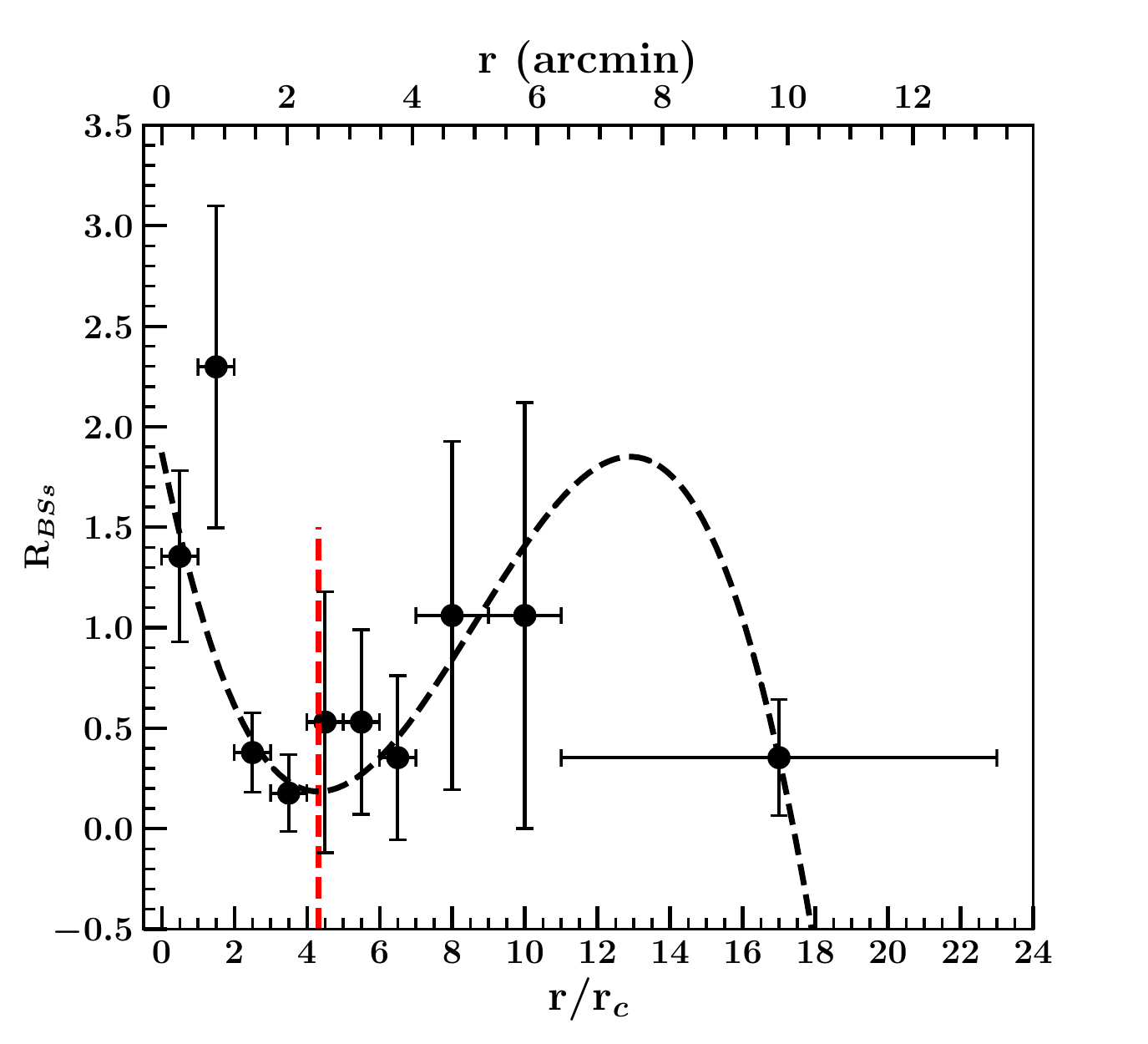}
    \includegraphics[width=0.49\textwidth]{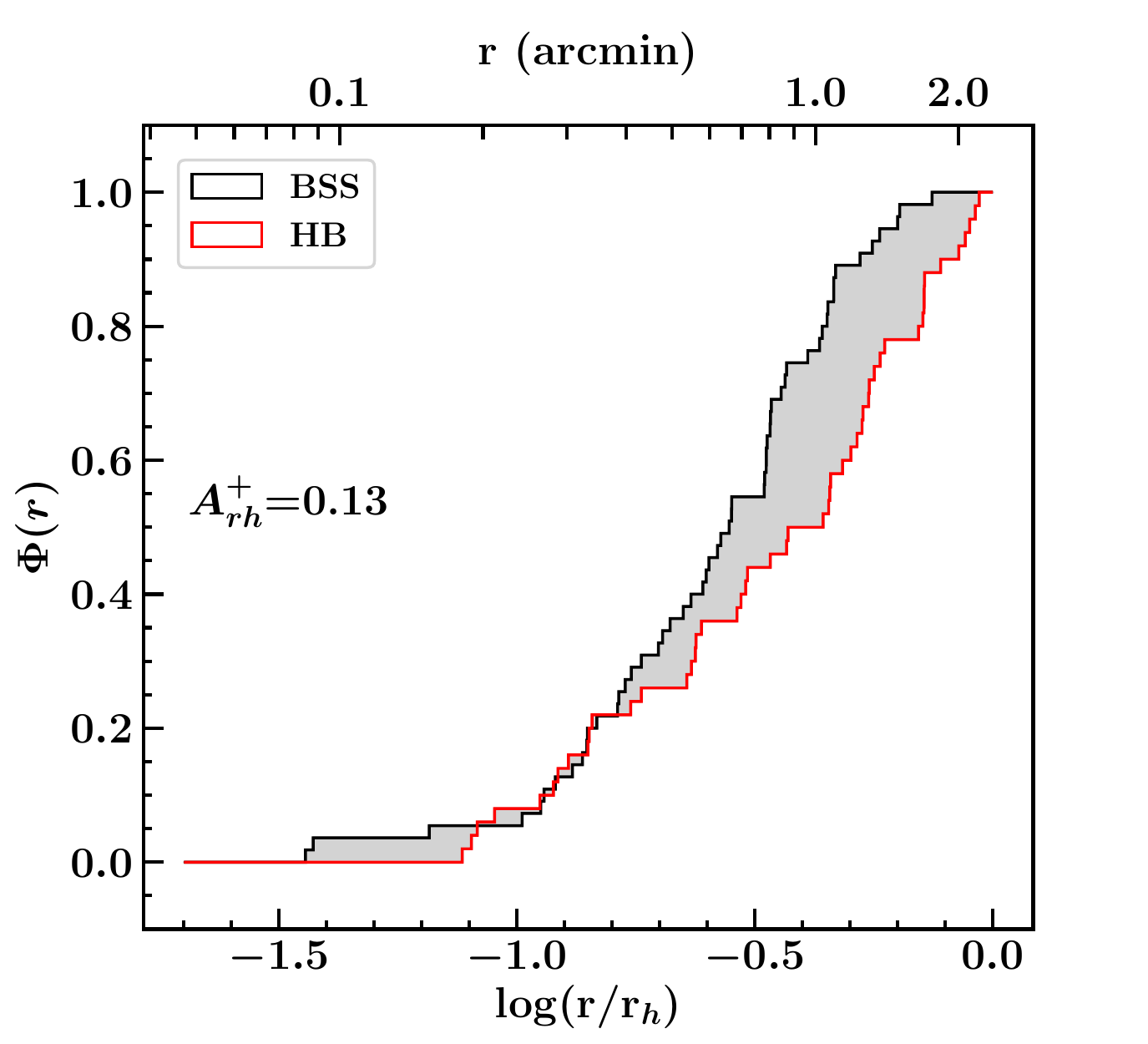}
    \caption{Left panel: normalized cumulative radial distribution of BSs up to 23 r$_c$. The r$_c$ value for the cluster is 0.58$'$. The black solids are calculated BSs-nRD at radial bins in unit of r$_c$. The dashed line is weighted 3rd order polynomial fit on the observed BSs-nRD. The red dashed line represents the minima of polynomial at a radial distance of 4.3 r$_c$ (2.5$'$) from the cluster center. Right panel: cumulative distribution of BSs and HB stars upto half-mass radius (2.358$'$) of the cluster. The black and red lines are cumulative radial distribution of BSs and HB stars, respectively. The gray area represents the difference in their radial distributions.}
    \label{fig:cum_BSs}
\end{figure*}

The BSs are good tracer of the dynamical status of the cluster \citep{Ferraro2012}. Based upon the relative radial distribution of BSs, \citet{Ferraro2012} suggested three groups of among GGCs: (i) Family I: dynamically young GCs show a flat radial distribution of BSs; (ii) Family II: dynamically intermediate GCs show a minima at intermediate radii from the cluster center; (iii) Family III: dynamically old/relaxed GCs show a central peak and a decreasing trend outward. The radial distribution of BSs is generally compared with either total luminosity distribution of the cluster or with any one of the evolutionary phases (MS, RGB, or HB). Here we will compare the radial distribution of BSs with HB stars to study the dynamical status of the cluster as HB stars are brighter than other evolutionary phases in UV photometry of the cluster. 

Our observed BSs distribution extends up to tidal radius of the cluster. Since the observed BSs population is not complete towards the cluster core, we adopted \citet{Simunovic2016} BSs for the {\em HST} observed region along with the additional BSs identified in this paper. We finally acquired a total of 66 BSs up to the tidal radius of the cluster. We were able to extract all the HB stars of the cluster using UVIT observations except one in the core which was blended with a BS. We included this HB star in our radial distribution based upon its {\em HST} observed position. We included BHBs, RHBs, and the newly identified EHB star and acquired a final HB catalog of 70 stars up to the tidal radius of the cluster.  

We calculated BSs normalised radial distribution (BSs-nRD) using \citet{Ferraro2012} prescription
\begin{equation}
\centering
    R_{BSs}(r)=\frac{N_{BSs}(r)/N_{BSs,\ tot}}{N_{HB}(r)/N_{HB,\ tot}}
    \label{eq:RBSs}
\end{equation}

\noindent where, $N_{BSs}(r)$ ($N_{HB}(r)$) is the number of BSs (HB stars) in r$^{th}$ radial bin and $N_{BSs,tot}$ ($N_{HB,tot}$) is the total number of BSs (HB stars) observed up to tidal radius of the cluster. The BSs-nRD is plotted in the left panel of \autoref{fig:cum_BSs}. The radial bins were converted in the core radius unit (r$_c$ = 0.58$'$). We have chosen the bin sizes such that each bin must contain at least one BS star and one HB star, so that the normalized ratio will not be zero. Upto 7r$_c$, the bin sizes are equal and each bin size is one r$_c$. Beyond 7r$_c$, the bin sizes are unequal. We have shown the bin sizes with horizontal lines in the left panel of \autoref{fig:cum_BSs}. The error bars were calculated using Poisson noise on the number counts of BSs and HB stars in each radial bin and following error propagation law. We can see a peak in the central region up to 2r$_c$ and then a minima after that around 3-4r$_c$ followed by an increasing trend in the BSs-nRD. We fitted a weighted 3rd order polynomial on the observed R$_{BSs}$(r) with weight as 1/(error in R$_{BSs}$). The fitted polynomial is shown in the dashed line in the figure. Using the fitted polynomial we estimated r$_{minima}$ at 4.3r$_c$ or at 2.5 arcmin.

In the right panel of \autoref{fig:cum_BSs}, we have plotted the normalized radial distribution of of BSs and HB stars up to half-mass radius (r$_{h,m}$ = 2.358$'$) of the cluster. We calculated the A$^+_{rh}$ parameter using \citet{Alessandrini2016} and \citet{Lanzoni2016} prescription

\begin{equation}
\centering
    A^+(x)=\int^{x}_{x_{min}} (\phi_{BSs}(x') - \phi_{HB}(x'))\ dx'
    \label{eq:aplus}
\end{equation}
where $\phi$(x) is the normalised cumulative radial distribution of each population, $x=\log (r/r_h)$ is the logarithmic distance from the cluster center normalized to the half-mass radius, r$_h$ and x$_{min}$ is the minimum distance sampled. The half-mass radius (r$_{h,m}$ = 7.58 pc) was taken from the GGC catalog\footnote{\href{https://people.smp.uq.edu.au/HolgerBaumgardt/globular/}{https://people.smp.uq.edu.au/HolgerBaumgardt/globular/}} \citep{Hilker2020}. We converted the r$_{h,m}$ from parsec to arcmin using cluster distance 11 kpc which gives r$_{h,m}$=2.358$'$. We found A$^+_{rh}$ = +0.13 using \autoref{eq:aplus} on the normalised cumulative radial distribution plot (right panel of \autoref{fig:cum_BSs}).

The above derived r$_{minima}$ and A$^+_{rh}$ suggest that the cluster belongs to Familly II with intermediate dynamical age. The value of  r$_{minima}$ at 4.3 r$_c$ shows similar r$_{minima}$ as in M53 and NGC 288 \citep{Ferraro2012, Sahu2019} and dynamically youngest in the Family II clusters \citep{Ferraro2012}. The value A$^+_{rh}$ of this cluster also supports the same status as shown in the Figure 1 and Table 1 of \citet{Lanzoni2016}.

\citet{Ferraro2018} derived the A$^+_{rh}$ as +0.02 using BSs and MSTO population. They used half-light radius, r$_h$=1.51$'$ assuming it is similar to the half-mass radius. However, using N-body simulations of 165 GCs, \citet{Baumgardt2018} found that the 3D-projection of half-mass radius (r$_{h,m}$) is different from the projected half-light radius (r$_{h,l}$) and they estimated (r$_{h,m}$) as 7.71 pc (2.35$'$) and (r$_{h,l}$) as 4.51 pc (1.51$'$). We found similar A$^+_{rh}$ (+0.027) as in \citet{Ferraro2018} when using  r$_h$ as 1.51$'$. We calculated the current core relaxation time (t$_{rc}$) using equation 2 of \citet{Ferraro2018}
\begin{equation}
    \log N_{relax} = 5.1(\pm0.5)\times A^+_{rh} + 0.79(\pm0.12)
\end{equation}
where N$_{relax}$ = t$_{GC}$ / t$_{c}$ and the A$^+_{rh}$ are derived in this paper. We found the dynamical age of the cluster as $ t_{c}=0.423\pm0.096$ Gyr. 
 
\section{Conclusion}
\label{sec:conclusion}
 
This paper presents the UV study of the cluster NGC 4590 using three FUV and three NUV filters of UVIT. We list below the important findings from our analysis. 
 \begin{enumerate}
    \item The cluster memberships of the observed sources were confirmed using the {\em HST} proper motion catalog for the inner region and the {\em Gaia} EDR3 proper motion catalog for the outer region. Finally, we got a total of $1442$ sources in NUV and 134 sources in FUV as cluster members.  
    
    \item The evolutionary stages of various stellar objects were traced using UV-optical CMDs. The different kind of sources identified are: $57$ BHBs, $11$ RHBs, $246$ RGBs, $140$ SGBs, $254$ main-sequence turn-off stars, $626$ main-sequence stars,  $41$ RR Lyraes, $31$ BSs, two SXPhes and one AGB star. We also detected two new FUV bright sources in the core of the cluster. One of them is an EHB star and the other one is either a post-BHk or probable WD star. 
    
    \item The EHB star, BHBs, post-BHk/WD and BSs are detected both in FUV and NUV filters of UVIT whereas the rest of the sources are detected only in NUV but not in FUV. So, the former sources are the main contributors to the FUV emission of the cluster.
    
    \item The physical parameters of the UV bright  sources were derived by fitting SEDs %(Kurucz models and Lavenhagen WD model) 
    to the observed photometric magnitudes. The range of T$_{\mathrm{eff}}$, L$_{\mathrm{bol}}$, radius and $\log$(g) of BHBs is $7,250 - 10,500$ K, $45.14 - 62.79$ L$_\odot$, $2.06 - 4.99$ R$_\odot$, and $1.5 - 3.5$ dex, respectively. Similarly, the range of T$_{\mathrm{eff}}$, L$_{\mathrm{bol}}$, radius and $\log$(g) of RHBs is $\sim6,000$ K, $69.88 - 86.88$ L$_\odot$, $7.72 - 10.42$ R$_\odot$, and $1.5 - 2.5$ dex, respectively .
    
    \item Using suitable HB evolutionary tracks, the range of masses of BHBs and RHBs obtained are $0.60 - 0.65$ M$_\odot$ and $0.65 - 0.70$ M$_\odot$, respectively. 
     
    \item We derived the physical parameters of the newly identified post-HBk/WD (in UV-optical CMD) using the SED fittings. The observed fluxes of the source was fitted with both the Kurucz model and the Levenhagen WD. The difference in the reduced chi-sqaure values between the two best fitted grids is very small. Hence, we chose the average values of effective temperature and log(g) derived from both the models. The average values are T$_{\mathrm{eff}}$ = 57,500$\pm$7,500 K and $\log$(g) = 6.0$\pm$1.0 dex, respectively. The BHk evolutionary tracks on T$_{\mathrm{eff}}$ vs. $\log$(g) plane suggest that the star is in the post-BHk phase lying in between early and late hot flash scenario post-BHk tracks. However, the T$_{\mathrm{eff}}$ vs. L$_{\mathrm{bol}}$ relation suggest that the star is nearing to complete its post-BHk evolutionary phase and about to begin its WD cooling phase. Hence, a high resolution UV/optical spectroscopy is needed to clearly probe its evolutionary phase.
     
    \item The derived T$_{\mathrm{eff}}$, L$_{\mathrm{bol}}$, radius and $\log$(g) of the BSs are in the range 5,750 $-$ 9,000 K, 3.12 $-$ 24.25 L$_\odot$, 0.93 $-$ 3.76 R$_\odot$, and 1.5 $-$ 4.5 dex, respectively. We also estimated their masses and ages by isochrone fitting on the T$_{\mathrm{eff}}$ vs. L$_{\mathrm{bol}}$ plane which are found to be in the range of 0.84 to 1.26 M$_\odot$ and  1.5 to 10.0 Gyr, respectively. The BSs show two groups in the T$_{\mathrm{eff}}$ vs. L$_{\mathrm{bol}}$ plot: one towards bluer/hotter end with an average age of 3$-$6 Gyr and the other one towards redder/cooler end with an average age of 6$-$10 Gyr. This suggests the presence of both the formation scenarios (collision and mass-transfer) of BSs in the cluster. We also find one E-BS star with 2 Gyr age in its SGB evolutionary phase. Generally, most of the massive BSs are found in the MS/MSTO phase whereas the less massive BSs are evolved towards their SGB phase. The radial distribution of BSs suggests that massive and young BSs are situated at the core of the cluster whereas the older and less massive BSs are distributed over the entire cluster.
     
    \item The normalised radial distribution of BSs suggests a bi-modal distribution with r$_{minima}$ at 2.5$'$ (4.3 r$_c$). We calculated the A$^+_{rh} = +0.13$ for the cluster using half-mass radius as 7.58 pc. We found that the cluster belongs to Family II and is one of the youngest clusters among dynamically intermediate age GGCs. We calculated the dynamical age of the cluster to be $0.423\pm0.096$ Gyr.
     
 \end{enumerate}

\section*{Acknowledgements}
We thank the anonymous reviewer for several useful comments and suggestions, which greatly improved the scientific contents of the paper. We would like to thank Dr. Zhenxin Lei for providing BHk evolutionary tracks. RK would like to acknowledge CSIR Research Fellowship (JRF) Grant No. 09/983(0034)/2019-EMR-1 for the financial support. ACP would like to acknowledge the support by Indian Space Research Organization, Department of Space, Government of India (ISRO RESPOND project  No.ISRO/RES/2/409/17-18). ACP thanks Inter University centre for Astronomy and Astrophysics (IUCAA), Pune, India for providing facilities to carry out his work. DKO acknowledges the support of the Department of Atomic Energy, Government of India, under Project Identification No. RTI 4002.  SC acknowledges support from Progetto Mainstream INAF (PI: S. Cassisi), from INFN (Iniziativa specifica TAsP), and from PLATO ASI-INAF agreement n.2015-019-R.1-2018. SC warmly thanks the Instituto de Astrofisica de Canarias for the hospitality and the \lq{Programa de investigadores visitantes}\rq  de la Fundacion Jesus Serra. AM thanks DST-INSPIRE (IF150845) for the funding.

This publication uses the data from the \mbox{{\em AstroSat}} mission of the Indian Space Research  Organisation (ISRO), archived at the Indian Space Science Data Center (ISSDC). The UVIT data used here was processed by the Payload Operations Centre at IIA. The UVIT is built in collaboration between IIA, IUCAA, TIFR, ISRO and CSA.

\section*{DATA AVAILABILITY}
The data underlying this article will be shared on reasonable request to the corresponding author.
    
%%gro%%%%%%%%%%%%%%%%%%%%%%%%%%%%%%%%%%%%%%%%%%%%%%%%



%\bibliographystyle{mnras}

\bibliographystyle{mnras}
\bibliography{4590}

%%%%%%%%%%%%%%%%%%%%%%%%%%%%%%%%%%%%%%%%%%%%%%5%%%%Tables%%%%%%%%%%%%%%%%%%%%%%%%%%%%%%%%%%%%%%%%%%%%%%%%%%%%%%

\begin{table}
\centering
\caption{UVIT observation details of NGC 4590. N$_f$, N$_{phot}$, and  N$_{cl}$ represent the number of frames/orbits,  number of total observed sources, and  number of cluster member sources in each UVIT filter.}
\label{tab:observation}
\resizebox{\columnwidth}{!}{
\begin{tabular}{lccccccr}
\hline
Chanel & Filter &   mean $ \lambda $ & $\Delta \lambda$ & N$_{f}$ & Exp. Time & N$_{phot}$ & N$_{cl}$ \\
    &   & (\AA)            & (\AA)            &                        & (sec.)    &            &         \\
\hline
\multirow{3}{*}{NUV}
& NUVB4  & 2632            & 275              &  2           & 1707.00              & 5964      & 1426  \\
& NUVB13 & 2447            & 280              &  1           & 1739.64              & 5418      & 1315  \\
& NUVB15 & 2196            & 270              &  1           & 1221.06              & 848       & 231   \\
\hline
\multirow{3}{*}{FUV}
& Silica & 1717            & 125              &  2           & 1405.86              & 1096      & 134  \\
& Sapphire & 1608          & 290              &  1           & 342.15               & 173       & 86  \\
& BaF2   & 1541            & 380              &  2           & 860.62               & 276       & 103 \\
\hline
\end{tabular}
}
\end{table}

\begin{table}
    \caption{UVIT detected stars of different evolutionary phases are listed below. Third column contains the symbols and colors used to denote the different sources in CMD (\autoref{fig:uv-opt_cmd}). }
    \label{tab:evol_phase}
    \centering
    \begin{tabular}{l c c}
    \hline
        Evol. phase & N$_{stars}$ & Color and Shape \\
    \hline
         EHB & 1  & brown diamond \\
         BHB  & 57 & blue circles \\
         RHB   & 11 & blue upper triangles \\
         RGB   & 246 & red circles\\
         SGB   & 140 & orange lower triangles \\
         MSTO  & 254 & red lower triangles \\
         MS    & 626 & gray circles \\
         SXPhes & 2 & violet diamonds \\
         RR Lyares & 41 & dark orange diamonds\\
         BSs & 31 &  blue-violet solids \\
         AGB & 1 & green upper triangle\\
         post-BHk/WD & 1 & black diamond \\
    \hline
    \end{tabular}
\end{table}

\begin{table*}
    \caption{List of the telescopes and their filters used in the SEDs fittings of BSs, BHBs, RHBs, and EHBs. }
    \label{tab:telescope}
    \centering
    
    \adjustbox{max width=\textwidth} {
    \begin{tabular}{ c c c  c  }
    \hline
    Telescope & Filters & Wavelength range &  Reference\\
    \hline
    UVIT & BaF2, Sapphire, Silica, NUVB15, NUVB13, and NUVB4 & 1350-2800 \AA &  This paper \\ 
    {\em GALEX} & FUV, NUV & 1350-3000 \AA & \citet{Schiavon2012} \\
    {\em HST} & F275W, F336W, F438W, F606W, and F814W  & 2200-9600 \AA & \citet{Nardiello2018} \\
    {\em Gaia}  & G, BP, RP & 3300-10600 \AA &  \citet{GaiaCatalog2018} \\
    PAN-STARRS & g, r, i, z, y & 3900-10800 \AA & \citet{Chambers2016} \\
    CTIO-4m  & U, B, V, R, I & 3000-11800 \AA &  \citet{Stetson2019} \\
    2MASS  & J, H, K       &  12000-21500 \AA  &  \citet{Skrutskie2006} \\

    \hline
    \end{tabular} }
\end{table*}

\begin{table*}
    \centering
    \caption{Physical parameters of BHB stars obtained from the SED fitting. The typical errors are 250 K and 0.25 dex in all the derived T$_{\mathrm{eff}}$ and $\log(g)$ values, respectively.}
    \label{tab:BHB}
    \begin{tabular}{l c c c c c c r }
\hline
  \multicolumn{1}{c }{Object} &
  \multicolumn{1}{c }{RA} &
  \multicolumn{1}{c }{DEC} &
  \multicolumn{1}{c }{T$_{\mathrm{eff}}$} &
  \multicolumn{1}{c }{$\log(g)$} &
  \multicolumn{1}{c }{L$_{bol}$} &
  \multicolumn{1}{c }{Radius} &
  \multicolumn{1}{r }{Radial distance}\\

  \multicolumn{1}{c }{ } &
 \multicolumn{2}{c}{(J2000, degrees)} &
  \multicolumn{1}{c }{(K)} &
  \multicolumn{1}{c }{(dex)} &
  \multicolumn{1}{c }{(L$_\odot$)} &
  \multicolumn{1}{c }{(R$_\odot$)} &
  \multicolumn{1}{r }{(arc-seconds)}\\
\hline
  BHB01 & 189.87083 & -26.745861 & 7250 & 2.5 & 57.84$\pm$10.97 & 4.83$\pm$0.22 & 17.26\\
  BHB02 & 189.87666 & -26.750278 & 7500 & 2.5 & 57.87$\pm$07.31 & 4.50$\pm$0.20 & 41.83\\
  BHB03 & 189.88687 & -26.764528 & 7500 & 2.5 & 72.05$\pm$12.78 & 4.99$\pm$0.23 & 101.60\\
  BHB04 & 189.90604 & -26.741417 & 7500 & 2.5 & 59.35$\pm$09.60 & 4.62$\pm$0.21 & 126.57\\
  BHB05 & 189.84262 & -26.745056 & 7750 & 2.5 & 56.37$\pm$06.99 & 4.22$\pm$0.19 & 77.84\\
  BHB06 & 189.86667 & -26.738111 & 7750 & 2.5 & 59.10$\pm$07.56 & 4.31$\pm$0.20 & 16.88\\
  BHB07 & 189.84879 & -26.734833 & 7750 & 2.5 & 62.79$\pm$07.83 & 4.42$\pm$0.20 & 64.33\\
  BHB08 & 189.85338 & -26.734528 & 7750 & 2.5 & 55.78$\pm$06.74 & 4.20$\pm$0.19 & 52.16\\
  BHB09 & 189.85367 & -26.727777 & 7750 & 3.0 & 56.89$\pm$07.38 & 4.14$\pm$0.19 & 68.41\\
  BHB10 & 189.86980 & -26.764778 & 7750 & 2.5 & 53.46$\pm$06.54 & 4.07$\pm$0.18 & 79.75\\
  BHB11 & 189.86595 & -26.739860 & 7750 & 2.5 & 59.04$\pm$08.52 & 4.23$\pm$0.19 & 10.85\\
  BHB12 & 189.85762 & -26.739027 & 7750 & 2.5 & 61.13$\pm$08.17 & 4.37$\pm$0.20 & 32.20\\
  BHB13 & 189.86342 & -26.738056 & 7750 & 2.0 & 57.46$\pm$08.40 & 4.18$\pm$0.19 & 20.07\\
  BHB14 & 189.91454 & -26.732555 & 7750 & 2.5 & 57.80$\pm$10.09 & 4.25$\pm$0.19 & 158.17\\
  BHB15 & 189.87200 & -26.714945 & 7750 & 3.0 & 57.71$\pm$10.64 & 4.16$\pm$0.19 & 101.72\\
  BHB16 & 189.93216 & -26.713667 & 7750 & 2.5 & 54.70$\pm$09.23 & 4.14$\pm$0.19 & 235.16\\
  BHB17 & 189.86258 & -26.746944 & 8000 & 2.0 & 54.60$\pm$06.62 & 3.86$\pm$0.18 & 19.95\\
  BHB18 & 189.87738 & -26.743917 & 8000 & 3.0 & 54.16$\pm$06.94 & 3.83$\pm$0.17 & 34.57\\
  BHB19 & 189.86400 & -26.744833 & 8000 & 1.0 & 55.73$\pm$07.20 & 3.91$\pm$0.18 & 11.35\\
  BHB20 & 189.87580 & -26.747360 & 8000 & 2.5 & 54.85$\pm$07.10 & 3.86$\pm$0.18 & 33.55\\
  BHB21 & 189.84187 & -26.759972 & 8000 & 2.5 & 58.81$\pm$09.85 & 3.99$\pm$0.18 & 100.96\\
  BHB22 & 189.93184 & -26.725666 & 8000 & 2.5 & 55.85$\pm$09.44 & 3.92$\pm$0.18 & 218.33\\
  BHB23 & 189.84096 & -26.714056 & 8000 & 2.5 & 55.42$\pm$09.81 & 3.86$\pm$0.18 & 132.50\\
  BHB24 & 189.85988 & -26.745861 & 8250 & 1.5 & 59.51$\pm$07.36 & 3.78$\pm$0.17 & 24.54\\
  BHB25 & 189.87650 & -26.751000 & 8250 & 2.5 & 55.17$\pm$06.70 & 3.63$\pm$0.16 & 43.18\\
  BHB26 & 189.87750 & -26.749750 & 8250 & 2.5 & 53.92$\pm$06.31 & 3.60$\pm$0.16 & 42.80\\
  BHB27 & 189.86325 & -26.739666 & 8250 & 1.5 & 55.51$\pm$06.77 & 3.61$\pm$0.16 & 15.82\\
  BHB28 & 189.87305 & -26.749971 & 8250 & 1.0 & 55.97$\pm$06.64 & 3.66$\pm$0.17 & 32.91\\
  BHB29 & 189.84267 & -26.768333 & 8250 & 2.5 & 54.96$\pm$09.17 & 3.65$\pm$0.17 & 120.07\\
  BHB30 & 189.73962 & -26.824750 & 8250 & 3.0 & 55.16$\pm$08.96 & 3.62$\pm$0.16 & 503.82\\
  BHB31 & 189.93463 & -26.752861 & 8250 & 2.5 & 54.98$\pm$09.32 & 3.62$\pm$0.16 & 221.36\\
  BHB32 & 189.85446 & -26.756805 & 8500 & 1.5 & 53.30$\pm$06.42 & 3.35$\pm$0.15 & 63.96\\
  BHB33 & 189.87633 & -26.723028 & 8500 & 3.0 & 52.88$\pm$06.27 & 3.34$\pm$0.15 & 77.62\\
  BHB34 & 189.85250 & -26.762362 & 8500 & 2.5 & 52.70$\pm$06.18 & 3.35$\pm$0.15 & 83.92\\
  BHB35 & 189.87671 & -26.761778 & 8500 & 3.0 & 51.55$\pm$06.42 & 3.30$\pm$0.15 & 75.52\\
  BHB36 & 189.86270 & -26.751444 & 8500 & 2.5 & 52.82$\pm$06.40 & 3.36$\pm$0.15 & 33.67\\
  BHB37 & 189.83930 & -26.957361 & 8500 & 3.0 & 53.92$\pm$08.78 & 3.38$\pm$0.15 & 777.42\\
  BHB38 & 189.82726 & -26.781973 & 8500 & 2.5 & 53.40$\pm$06.86 & 3.36$\pm$0.15 & 189.63\\
  BHB39 & 189.86433 & -26.793500 & 8500 & 3.0 & 53.60$\pm$08.77 & 3.36$\pm$0.15 & 182.68\\
  BHB40 & 189.86826 & -26.748278 & 8750 & 2.5 & 50.59$\pm$05.94 & 3.10$\pm$0.14 & 20.35\\
  BHB41 & 189.86671 & -26.739555 & 8750 & 2.5 & 60.24$\pm$07.28 & 3.35$\pm$0.15 & 11.68\\
  BHB42 & 189.83713 & -26.735306 & 8750 & 2.0 & 52.00$\pm$06.41 & 3.13$\pm$0.14 & 98.82\\
  BHB43 & 189.83446 & -26.753000 & 8750 & 3.0 & 58.13$\pm$09.39 & 3.31$\pm$0.15 & 109.96\\
  BHB44 & 189.88737 & -26.782278 & 8750 & 2.0 & 55.21$\pm$07.00 & 3.20$\pm$0.15 & 156.88\\
  BHB45 & 189.84592 & -26.638500 & 8750 & 2.0 & 53.53$\pm$08.69 & 3.16$\pm$0.14 & 381.38\\
  BHB46 & 189.85037 & -26.743305 & 9000 & 3.0 & 49.69$\pm$05.81 & 2.89$\pm$0.13 & 52.53\\
  BHB47 & 189.76150 & -26.654500 & 9000 & 3.5 & 49.42$\pm$08.30 & 2.87$\pm$0.13 & 464.24\\
  BHB48 & 189.89304 & -26.633806 & 9000 & 3.0 & 50.16$\pm$10.92 & 2.91$\pm$0.13 & 401.42\\
  BHB49 & 189.84483 & -26.748638 & 9250 & 2.5 & 53.55$\pm$06.01 & 2.83$\pm$0.13 & 73.38\\
  BHB50 & 189.86021 & -26.747028 & 9750 & 3.0 & 47.08$\pm$05.47 & 2.38$\pm$0.11 & 25.83\\
  BHB51 & 189.87038 & -26.744083 & 9750 & 3.0 & 47.41$\pm$05.61 & 2.38$\pm$0.11 & 12.70\\
  BHB52 & 189.87550 & -26.734540 & 10000 & 2.5 & 45.14$\pm$06.93 & 2.20$\pm$0.10 & 41.04\\
  BHB53 & 189.91490 & -26.842320 & 10500 & 3.0 & 48.75$\pm$07.43 & 2.09$\pm$0.09 & 390.32\\
  BHB54 & 189.83190 & -26.813600 & 10500 & 3.0 & 47.66$\pm$07.23 & 2.06$\pm$0.09 & 278.34\\
  BHB55 & 189.90590 & -26.751730 & 10250 & 3.5 & 45.28$\pm$07.37 & 2.10$\pm$0.10 & 130.06\\
  BHB56 & 189.82820 & -26.744210 & 10250 & 3.0 & 45.35$\pm$08.93 & 2.11$\pm$0.10 & 123.88\\
  BHB57 & 189.83930 & -26.756790 & 10000 & 2.5 & 47.88$\pm$03.99 & 2.26$\pm$0.00 & 101.46\\
\hline\end{tabular}

\end{table*}

\begin{table*}
    \centering
    \caption{Physical parameters of RHB stars obtained from the SED fitting.}
    \label{tab:RHB}
    \begin{tabular}{l c c c c c c r}
\hline
  \multicolumn{1}{c }{Object} &
  \multicolumn{1}{c }{RA} &
  \multicolumn{1}{c }{DEC} &
  \multicolumn{1}{c }{T$_{\mathrm{eff}}$} &
  \multicolumn{1}{c }{$\log(g)$} &
  \multicolumn{1}{c }{L$_{bol}$} &
  \multicolumn{1}{c }{Radius} &
  \multicolumn{1}{r }{Radial distance}\\

  \multicolumn{1}{c }{ } &
  \multicolumn{2}{c}{(J2000, degrees)} &
  \multicolumn{1}{c }{(K)} &
  \multicolumn{1}{c }{(dex)} &
  \multicolumn{1}{c }{(L$_\odot$)} &
  \multicolumn{1}{c }{(R$_\odot$)}&
  \multicolumn{1}{r }{(arc-seconds)}\\  
\hline
  RHB01 & 189.88583 & -26.757889 & 5750 & 2.5 & 69.88$\pm$11.03 & 8.31$\pm$0.38 & 82.05\\
  RHB02 & 189.87837 & -26.728195 & 5500 & 2.0 & 86.88$\pm$14.76 & 10.42$\pm$0.47 & 64.59\\
  RHB03 & 189.88579 & -26.739860 & 5750 & 2.5 & 75.73$\pm$07.81 & 8.83$\pm$0.40 & 62.28\\
  RHB04 & 189.87100 & -26.739529 & 5500 & 2.5 & 82.31$\pm$13.73 & 10.06$\pm$0.46 & 18.16\\
  RHB05 & 189.85683 & -26.752890 & 6000 & 2.0 & 72.12$\pm$10.53 & 7.94$\pm$0.36 & 48.23\\
  RHB06 & 189.92267 & -26.678389 & 6000 & 2.0 & 69.10$\pm$10.95 & 7.72$\pm$0.35 & 293.54\\
  RHB07 & 189.79813 & -26.795416 & 5750 & 2.5 & 72.65$\pm$10.82 & 8.56$\pm$0.39 & 290.61\\
  RHB08 & 189.88696 & -26.645695 & 6000 & 1.5 & 80.71$\pm$10.37 & 8.36$\pm$0.38 & 355.60\\
  RHB09 & 189.81421 & -26.775972 & 5750 & 2.0 & 77.20$\pm$09.84 & 8.88$\pm$0.40 & 206.71\\
  RHB10 & 189.86366 & -26.578861 & 5750 & 2.5 & 78.15$\pm$10.04 & 8.89$\pm$0.40 & 590.26\\
  RHB11 & 189.91376 & -26.770472 & 5750 & 2.0 & 74.18$\pm$09.51 & 8.69$\pm$0.39 & 181.13\\
\hline\end{tabular}

\end{table*}

\begin{table*}
    \centering
    \caption{Physical parameters of BSs obtained from the SED fitting. The age and mass of BSs are derived from the isochrone fitting on the T$_{\mathrm{eff}}$-L$_{bol}$ plane as shown in \autoref{fig:teff_bs}.}
    \label{tab:BSs}
    \begin{tabular}{l c c c c c c c c r}
\hline
  \multicolumn{1}{c }{Object} &
  \multicolumn{1}{c }{RA} &
  \multicolumn{1}{c }{DEC} &
  \multicolumn{1}{c }{T$_{\mathrm{eff}}$} &
  \multicolumn{1}{c }{$\log(g)$} &
  \multicolumn{1}{c }{L$_{bol}$} &
  \multicolumn{1}{c }{Radius} &
  \multicolumn{1}{c }{Age} &
  \multicolumn{1}{c }{Mass} &
  \multicolumn{1}{r }{Radial distance}\\

  \multicolumn{1}{c }{ } &
 \multicolumn{2}{c}{(J2000, degrees)} &
  \multicolumn{1}{c }{(K)} &
  \multicolumn{1}{c }{(dex)} &
  \multicolumn{1}{c }{(L$_\odot$)} &
  \multicolumn{1}{c }{(R$_\odot$)} &
  \multicolumn{1}{c }{(Gyr)} &
  \multicolumn{1}{c }{(M$_\odot$)}&
  \multicolumn{1}{r }{(arc-seconds)}\\  
\hline

  BS01 & 189.85600 & -26.751667 & 9000 & 4.0 & 5.55$\pm$0.26 & 0.93$\pm$0.04  & 1.50 & 1.18 & 46.93 \\
  BS02 & 189.86320 & -26.734583 & 8500 & 3.5 & 7.71$\pm$0.39 & 1.22$\pm$0.05  & 4.25 & 1.05 & 31.65\\
  BS03 & 189.88121 & -26.739445 & 8250 & 3.5 & 7.37$\pm$0.41 & 1.28$\pm$0.05  & 4.75 & 1.01 & 48.19\\
  BS04 & 189.91470 & -26.729445 & 8250 & 4.0 & 9.80$\pm$1.04 & 1.50$\pm$0.09  & 4.50 & 1.05 & 161.64\\
  BS05 & 189.94771 & -26.758389 & 8000 & 3.5 & 7.78$\pm$0.62 & 1.43$\pm$0.07  & 5.25 & 1.00 & 266.40\\
  BS06 & 189.89317 & -26.788110 & 8000 & 4.0 & 4.78$\pm$0.41 & 1.12$\pm$0.06  & 5.00 & 0.98 & 183.97\\
  BS07 & 189.89462 & -26.740223 & 7750 & 4.0 & 4.49$\pm$0.51 & 1.13$\pm$0.08  & 6.00 & 0.94 & 90.24\\
  BS08 & 189.83795 & -26.756945 & 8750 & 4.0 & 13.88$\pm$0.68 & 1.62$\pm$0.06 & 3.25 & 1.12 & 105.52\\
  BS09 & 189.87800 & -26.738472 & 8750 & 4.0 & 8.53$\pm$0.55 & 1.26$\pm$0.05  & 3.75 & 1.08 & 39.53\\
  BS10 & 189.86750 & -26.733000 & 8750 & 4.5 & 6.94$\pm$0.48 & 1.12$\pm$0.05  & 3.50 & 1.08 & 35.37\\
  BS11 & 189.86888 & -26.752556 & 8250 & 3.5 & 13.92$\pm$0.65 & 1.82$\pm$0.07 & 3.50 & 1.13 & 35.81\\
  BS12 & 189.87009 & -26.735445 & 9500 & 3.0 & 16.83$\pm$0.57 & 1.49$\pm$0.05 & 2.50 & 1.24 & 28.63\\
  BS13 & 189.85805 & -26.731833 & 8500 & 4.0 & 8.69$\pm$0.66 & 1.35$\pm$0.07  & 4.25 & 1.05 & 48.29\\
  BS14 & 189.84654 & -26.745583 & 8000 & 3.0 & 9.30$\pm$0.61 & 1.58$\pm$0.07  & 4.75 & 1.03 & 65.58\\
  BS15 & 189.85571 & -26.752945 & 8250 & 3.0 & 6.96$\pm$0.25 & 1.29$\pm$0.05  & 4.75 & 1.01 & 50.81\\
  BS16 & 189.85487 & -26.746195 & 7000 & 2.5 & 24.25$\pm$0.81 & 3.34$\pm$0.13 & 2.50 & 1.26 & 39.95\\
  BS17 & 189.86371 & -26.736944 & 7750 & 2.5 & 13.43$\pm$0.55 & 2.03$\pm$0.08 & 4.00 & 1.09 & 23.17\\
  BS18 & 189.87038 & -26.727083 & 8250 & 3.5 & 6.93$\pm$0.28 & 1.28$\pm$0.05  & 4.75 & 1.01 & 57.80\\
  BS19 & 189.88254 & -26.760610 & 8500 & 3.5 & 7.00$\pm$0.18 & 1.22$\pm$0.04  & 4.25 & 1.04 & 81.88\\
  BS20 & 189.87675 & -26.744473 & 8250 & 3.5 & 10.70$\pm$0.45 & 1.58$\pm$0.06 & 4.25 & 1.07 & 32.87\\
  BS21 & 189.87267 & -26.736500 & 7750 & 3.5 & 9.09$\pm$0.43 & 1.66$\pm$0.07  & 5.25 & 1.00 & 29.71\\
  BS22 & 189.86626 & -26.741388 & 8000 & 2.5 & 9.14$\pm$0.51 & 1.58$\pm$0.07  & 4.75 & 1.03 & 5.28\\
  BS23 & 189.84541 & -26.693695 & 6000 & 3.0 & 10.10$\pm$0.70 & 2.94$\pm$0.16 & 6.50 & 0.95 & 189.57\\
  BS24 & 189.85371 & -26.676306 & 5750 & 3.0 & 13.89$\pm$1.35 & 3.76$\pm$0.24 & 4.75 & 1.05 & 189.57\\
  BS25 & 189.81505 & -26.795140 & 6000 & 3.5 & 7.53$\pm$0.26 & 2.54$\pm$0.11  & 8.75 & 0.87 & 251.12\\
  BS26 & 189.90196 & -26.668222 & 7250 & 4.0 & 3.12$\pm$0.14 & 1.12$\pm$0.05  & 7.50 & 0.87 & 291.44\\
  BS27 & 189.78633 & -26.796055 & 6750 & 3.0 & 4.99$\pm$0.21 & 1.63$\pm$0.07  & 9.75 & 0.84 & 321.69\\
  BS28 & 189.86842 & -26.952194 & 6500 & 3.5 & 5.46$\pm$0.12 & 1.85$\pm$0.07  & 10.0 & 0.84 & 753.84\\
  BS29 & 190.09113 & -26.696527 & 6500 & 1.5 & 5.32$\pm$0.18 & 1.82$\pm$0.08  & 10.0 & 0.84 & 740.65\\
  BS30 & 189.85100 & -26.642056 & 6250 & 2.5 & 8.28$\pm$0.13 & 2.46$\pm$0.10  & 7.50 & 0.91 & 366.18\\
  BS31 & 189.87013 & -26.755444 & 8000 & 4.0 & 5.17$\pm$0.11 & 1.19$\pm$0.04  & 5.25 & 0.97 & 46.84\\
\hline\end{tabular}

\end{table*}

\begin{table*}
    \centering
    \caption{Physical parameters of EHB, post-BHk/WD, and AGB stars obtained from the SED fitting.}
    \label{tab:misc}
     \begin{tabular}{l c c c c c c r}
\hline
  \multicolumn{1}{c }{Object} &
  \multicolumn{1}{c }{RA} &
  \multicolumn{1}{c }{DEC} &
  \multicolumn{1}{c }{T$_{\mathrm{eff}}$} &
  \multicolumn{1}{c }{$\log(g)$} &
  \multicolumn{1}{c }{L$_{bol}$} &
  \multicolumn{1}{c }{Radius} &
  \multicolumn{1}{r }{Radial distance}\\

  \multicolumn{1}{c }{ } &
  \multicolumn{2}{c}{(J2000, degrees)} &
  \multicolumn{1}{c }{(K)} &
  \multicolumn{1}{c }{(dex)} &
  \multicolumn{1}{c }{(L$_\odot$)} &
  \multicolumn{1}{c }{(R$_\odot$)} &
  \multicolumn{1}{r }{(arc-seconds)}\\  
\hline
  AGB & 189.874375 & -26.739860 & 5,500 & 2.0 & 176.5$\pm$16.56 & 14.71$\pm$0.67 & 26.85\\
  EHB & 189.874375 & -26.761361 & 28,000 & 5.0 & 14.52$\pm$0.36 & 0.161$\pm$0.006 & 71.23\\
  post-BHk/WD & 189.852542 & -26.759667 & 57,500$\pm$7,500 & 6.0$\pm$1.0 & 11.50$\pm$3.45 & 0.035$\pm$0.015 & 75.88\\
\hline\end{tabular}

\end{table*}

%%%%%%%%%%%%%%%%%%%%%%%%%%%%%%%%%%%%%%%%%%%%%%%%%% Appendix %%%%%%%%%%%%%%%%%%%%%%%%%%%%%%%%%%%%%%%%%%%%%%%%%%%%

%%%%%%%%%%%%%%%%% APPENDICES %%%%%%%%%%%%%%%%%%%%%
% Don't change these lines
\bsp	% typesetting comment
\label{lastpage}
\end{document}